\documentclass{aastex63}

\usepackage{multirow}
\usepackage{amsmath}
\usepackage{longtable}

\newcommand{\epsE}{$\epsilon$ Eridani}
\newcommand{\Hipp}{\textit{Hipparcos}}
\newcommand{\Gaia}{\textit{Gaia}}
\newcommand{\orbitize}{\texttt{orbitize!}}
\published{October 6, 2021}
\shorttitle{Constraining the Orbit and Mass of $\epsilon$ Eri b}
\shortauthors{Llop-Sayson et al.}
\graphicspath{{./}{figures/}}

\begin{document}

\title{Constraining the Orbit and Mass of $\epsilon$ Eridani b with Radial Velocities, Hipparcos IAD-{\Gaia~DR2} Astrometry, and Multi-epoch Vortex Coronagraphy Upper Limits}

\correspondingauthor{Jorge~Llop-Sayson}
\email{jllopsay@caltech.edu}
\author[0000-0002-3414-784X]{Jorge~Llop-Sayson}
\affiliation{California Institute of Technology, 1200 E California Blvd., Pasadena, CA 91125}
\author{Jason~J.~Wang}
\altaffiliation{51 Pegasi b Fellow}
\affiliation{California Institute of Technology, 1200 E California Blvd., Pasadena, CA 91125}
\author{Jean-Baptiste~Ruffio}
\affiliation{California Institute of Technology, 1200 E California Blvd., Pasadena, CA 91125}
\author[0000-0002-8895-4735]{Dimitri~Mawet}
\affiliation{California Institute of Technology, 1200 E California Blvd., Pasadena, CA 91125}
\affiliation{Jet Propulsion Laboratory, California Institute of Technology, 4800 Oak Grove Drive, Pasadena, CA 91109}
\author{Sarah~Blunt}
\affiliation{California Institute of Technology, 1200 E California Blvd., Pasadena, CA 91125}
\author{Olivier~Absil}
\altaffiliation{F.R.S.-FNRS Research Associate}
\affiliation{Space sciences, Technologies \& Astrophysics Research (STAR) Institute
Université de Liège (ULiège)
19c allée du Six Août, 4000 Liège, Belgium}
\author{Charlotte~Bond}
\affiliation{Science and Technology Facilities Council (STFC),
Polaris House,
North Star Avenue,
Swindon,
SN2 1SZ, United Kingdom
}
\author{Casey Brinkman}
\affiliation{Institute for Astronomy, University of Hawai'i, 2680 Woodlawn Drive, Honolulu, HI 96822, USA}
\author[0000-0003-2649-2288]{Brendan P. Bowler}
\affiliation{Department of Astronomy, The University of Texas at Austin, Austin, TX 78712, USA}
\author{Michael~Bottom}
\affiliation{Institute for Astronomy, University of Hawai'i, 2680 Woodlawn Drive, Honolulu, HI 96822, USA}
\author[0000-0003-1125-2564]{Ashley Chontos}
\altaffiliation{NSF Graduate Research Fellow}
\affiliation{Institute for Astronomy, University of Hawai'i, 2680 Woodlawn Drive, Honolulu, HI 96822, USA}
\author[0000-0002-4297-5506]{Paul A.\ Dalba}
\altaffiliation{NSF Astronomy and Astrophysics Postdoctoral Fellow}
\affiliation{Department of Earth and Planetary Sciences, University of California Riverside, 900 University Ave, Riverside, CA 92521, USA}
\author{B.J.~Fulton}
\affiliation{California Institute of Technology, 1200 E California Blvd., Pasadena, CA 91125}
\author{Steven Giacalone}
\affiliation{501 Campbell Hall, University of California at Berkeley, Berkeley, CA 94720, USA}
\author{Michelle Hill}
\affiliation{Department of Earth and Planetary Sciences, University of California Riverside, 900 University Ave, Riverside, CA 92521, USA}
\author[0000-0001-8058-7443]{Lea~A.~Hirsch}
\affiliation{Kavli Institute for Particle Astrophysics and Cosmology, Stanford University, Stanford, CA 94305, USA}
\author{Andrew~W.~Howard}
\affiliation{California Institute of Technology, 1200 E California Blvd., Pasadena, CA 91125}
\author[0000-0002-0531-1073]{Howard Isaacson}
\affiliation{501 Campbell Hall, University of California at Berkeley, Berkeley, CA 94720, USA}
\affiliation{Centre for Astrophysics, University of Southern Queensland, Toowoomba, QLD, Australia}
\author{Mikael~Karlsson}
\affiliation{Department of Materials Science and Engineering, Ångström Laboratory, Uppsala University, Uppsala, Sweden}
\author[0000-0001-8342-7736]{Jack Lubin}
\affiliation{Department of Physics \& Astronomy, University of California Irvine, Irvine, CA 92697, USA}
\author{Alex Madurowicz}
\affiliation{Kavli Institute for Particle Astrophysics and Cosmology, Stanford University, Stanford, CA 94305, USA}
\author{Keith~Matthews}
\affiliation{California Institute of Technology, 1200 E California Blvd., Pasadena, CA 91125}
\author{Evan~Morris}
\affiliation{Department of Astronomy \& Astrophysics, University of California, Santa Cruz, CA 95064, USA}
\author{Marshall~Perrin}
\affiliation{Space Telescope Science Institute, 3700 San Martin Drive
Baltimore, MD 21218}
\author[0000-0003-1698-9696]{Bin Ren}
\affiliation{California Institute of Technology, 1200 E California Blvd., Pasadena, CA 91125}
\author[0000-0002-7670-670X]{Malena Rice}
\affiliation{Department of Astronomy, Yale University, New Haven, CT 06511, USA}
\author{Lee~J.~Rosenthal}
\affiliation{California Institute of Technology, 1200 E California Blvd., Pasadena, CA 91125}
\author[0000-0003-4769-1665]{Garreth~Ruane}
\affiliation{Jet Propulsion Laboratory, California Institute of Technology, 4800 Oak Grove Drive, Pasadena, CA 91109}
\author{Ryan Rubenzahl}
\affiliation{California Institute of Technology, 1200 E California Blvd., Pasadena, CA 91125}
\author{He~Sun}
\affiliation{California Institute of Technology, 1200 E California Blvd., Pasadena, CA 91125}
\author{Nicole~Wallack}
\affiliation{California Institute of Technology, 1200 E California Blvd., Pasadena, CA 91125}
\author{Jerry~W.~Xuan}
\affiliation{California Institute of Technology, 1200 E California Blvd., Pasadena, CA 91125}
\author{Marie~Ygouf}
\affiliation{Jet Propulsion Laboratory, California Institute of Technology, 4800 Oak Grove Drive, Pasadena, CA 91109}

\begin{abstract}
\epsE~is a young planetary system hosting a complex multi-belt debris disk and a confirmed Jupiter-like planet orbiting at 3.48 AU from its host star. Its age and architecture are thus reminiscent of the early Solar System. The most recent study of \citet{Mawet2019}, which combined radial velocity (RV) data and Ms-band direct imaging upper limits, started to constrain the planet's orbital parameters and mass, but are still affected by large error bars and degeneracies. Here we make use of the most recent data compilation from three different techniques to further refine \epsE~b's properties: RVs, absolute astrometry measurements from the \Hipp~and \Gaia~missions, and new Keck/NIRC2 Ms-band vortex coronagraph images. We combine this data in a Bayesian framework. We find a new mass, $M_b$ = $0.66_{-0.09}^{+0.12}$~M$_{Jup}$, and inclination, $i$ = $78.81_{-22.41}^{\circ+29.34}$, with at least a factor 2 improvement over previous uncertainties. We also report updated 
constraints on the longitude of the ascending node, the argument of the periastron, and the time of periastron passage. With these updated parameters, we can better predict the position of the planet at any past and future epoch, which can greatly help define the strategy and planning of future observations and with subsequent data analysis. In particular, these results can assist the search for a direct detection with JWST and the Nancy Grace Roman Space Telescope's coronagraph instrument (CGI).

\end{abstract}


\section{Introduction}\label{sec:intro}
\epsE~is a nearby K2V dwarf star (Table \ref{table1}) surrounded by a prominent multi-belt debris disk and a confirmed Jupiter-like planet on a 7.4-year orbit \citep{Mawet2019}. Its relative young age, spectral type, and architecture are reminiscent of the early Solar System. Its proximity (3.2 pc) and thus apparent brightness ($V=3.73$ mag.) make it an excellent laboratory to study the early formation and evolution of planetary systems analogous to the Solar System.

Since the discovery of its debris disk by the Infrared Astronomical Satellite \citep[IRAS, ][]{Aumann1985}, \epsE~has been the subject of increased scrutiny culminating with the discovery of decade-long radial velocity (RV) variations by \citet{Hatzes2000} pointing towards the presence of a $\simeq 1.5$ M$_J$ giant planet with a period $P=6.9$ yr ($\simeq 3$ AU orbit) and a high eccentricity ($e=0.6$). This early orbital solution was dynamically incompatible with the multi-belt disk configuration of \epsE, raising the possibility of false alarm or at the very least confusion due to stellar jitter. {\citet{Benedict2006} attempted at modelling the perturbation caused by the companion using RV combined with HST astrometry.} Recently, \citet{Mawet2019} revisited this poster-child system by combining three decades of RV measurements with the most sensitive direct imaging data set ever obtained. The RV data, its state-of-the-art treatment of stellar noise, and direct imaging upper limits were combined in an innovative joint Bayesian analysis, providing new constraints on the mass and orbital parameters of the elusive planet. \citet{Mawet2019} reported a mass of $0.78^{+0.38}_{-0.12}$ $M_{Jup}$, semi-major axis (SMA) of $3.48\pm 0.02$ AU with a period of $7.37 \pm 0.07$ years, and an eccentricity of $0.07^{+0.06}_{-0.05}$, an order of magnitude smaller than earlier estimates and consistent with a circular orbit. These new orbital parameters were found to be dynamically compatible to the most recent picture of \epsE's disk architecture \citep{Su2017, Booth2017, Ertel2018}. However, the joint RV-direct imaging upper limit analysis of \citet{Mawet2019} left some orbital parameters such as inclination $i$, longitude of ascending node $\Omega$, and argument of periapsis $\omega$, largely unconstrained, making future pointed direct detection attempts (e.g. with JWST) more difficult. 

In this paper, we compiled data from recent \Gaia~data release 2 (DR2) forming a 25-year time baseline with archival \Hipp~astrometric data, plus new RV and deep direct imaging observations, to perform a joint astrometry-RV-direct imaging analysis aimed at refining the orbital elements of \epsE~b. The paper is organized as follows. In Sec.~\ref{sec:obs} we present the observations, namely, we include absolute astrometry measurements of \epsE~to help constrain the orbital parameters and mass. We include an updated set of RV measurements, and new direct imaging observations with Keck/NIRC2. Sec.~\ref{sec:analysis} describes the methods used to constrain the orbit and mass of \epsE~by combining the three different observation techniques. At the end of this section, the new constraints are presented. In Sec.~\ref{sec:discussion} we discuss the findings, update the planet-disk interaction parameters with respect to \citet{Mawet2019} and discuss the prospects of a detection with JWST.

\begin{deluxetable}{ccc}
\tabletypesize{\scriptsize}

\tablecaption{Properties of \epsE \label{table1}}
\tablewidth{0pt}
\tablehead{
\colhead{Property} & \colhead{Value} & \colhead{Ref.}}
\startdata
RA (hms) & 03 32 55.8 (J2000) &\citet{vanLeeuwen2007}\\
Dec (dms) & -09 27 29.7 (J2000) &\citet{vanLeeuwen2007}\\
Spectral type & K2V &\citet{Keenan1989}\\
Mass ($M_\odot$) & $0.82 \pm 0.02$ &\citet{Baines2012}\\
Distance (pc) & $3.216 \pm 0.0015$ &\citet{vanLeeuwen2007}\\
$V$ mag. & 3.73 &\citet{Ducati2002} \\ 
$K$ mag. & 1.67 &\citet{Ducati2002}\\
$L$ mag. & 1.60 &\citet{Cox2000}\\
$M$ mag. & 1.69 &\citet{Cox2000}\\
Age (Myr) &400-800 &\citet{Mamajek2008,Janson2015}\\
\enddata
\end{deluxetable}

%
% **********************************************************************
% **********************************************************************
% **********************************************************************
%

\section{Observations}\label{sec:obs}
\subsection{Radial Velocities}\label{sec:rv}

\epsE{} has been targeted by five multi-year radial velocity (RV) planet searches over the past 30 years. Most of the resulting RV measurements are presented and discussed in \citet{Mawet2019}. Since the publication of that paper, we have obtained 76 %(as of 11-17-20)
additional spectra of \epsE{} with Keck/HIRES \citep{Vogt1994}, and 172 %(as of 11-17-20)
spectra with the Levy Spectrograph on the Automated Planet Finder (APF) telescope  (\citealt{Radovan2014}; \citealt{Vogt2014}). These new measurements are given in Table \ref{tab:rvs}. As in \citet{Mawet2019}, new HIRES observations were taken using the standard iodine cell configuration used for the California Planet Survey \citep{Howard2010}. Observations were taken through either the B5 ($0.''87$ $\times$ $3.''5$) or C2 ($0.''87$ $\times$ $14.''0$) decker, yielding R~$\simeq$ 55,000 spectral resolution. Spectra had a median of 293,000 counts per exposure, corresponding to a median extracted flux of 67,000 counts (SNR = 260) at 550 nm. The same template spectrum described in \citet{Mawet2019}, taken in 2010 using the B3 decker, was used to derive RVs from all new spectra obtained for this work.

New APF measurements were also taken using the standard configuration described in \citet{Mawet2019}. Most APF measurements were acquired through the W decker ($1''$ x $3
''$), with a spectral resolution of R $\simeq$ 110,000. The median exposure time across observations was 26 s, yielding a per-pixel SNR of 120 at 550 nm. The same template spectrum described in \citet{Mawet2019}, taken in 2014 using the N decker, was used to derive RVs from all new spectra obtained for this work. For our analysis, all exposures taken with a single instrument within 10 hours were binned.

%
% **********************************************************************
%
\subsection{Absolute Astrometry Data}\label{sec:astrometry}
We obtain the astrometry data for \epsE~from the \Hipp~catalog Hip2 \citep{vanLeeuwen2007}, and the {Gaia DR2}~catalog \citep{Gaia2018}, as well as the intermediate astrometry data (IAD) available from the \Hipp~mission.  

% **********************************************************************
\subsubsection{\Hipp~Intermediate Astrometry Data}
The \Hipp~data is generally used in its reduced form, consisting of a five-element parameter set: the right-ascension $\alpha$, declination $\delta$, the proper motion (PM) vector $\mu_\alpha$ and $\mu_\delta$, and the parallax. However, the IAD contains all individual measurements taken by the mission, in the case of \epsE, from 1990.036 to 1992.563. The orbit period of \epsE~b being $\sim$7.31 years, \Hipp~IAD baseline is $\sim$0.35 times that period. Therefore, \Hipp~IAD may include changes in astrometry that contain signs of the presence of a companion. Indeed, a companion perturbs the otherwise constant rectilinear motion of the photocenter into a Keplerian motion around the barycenter. By using the IAD, we fit our model orbit to the curve described by the measured positions of \epsE's photocenter during the \Hipp~mission. 

The \Hipp~mission performed its measurements on a 1D scan, the acquisition protocol of which allowed for a reconstruction of the 1D points into 2D positions on the sky. The orbital direction of the scan, along with the precise epoch of each measurement are recorded in order to retrieve the final position. We use data from \cite{vanLeeuwen2007} that consists of the IAD residuals from the fit to the data \citep{vanLeeuwen2007}, the scan direction, epochs, and the errors associated with the original data. Using the fit from \cite{vanLeeuwen2007} and the residuals combined with the scan direction data, we reconstruct the one dimensional scan measurements following the method used in \cite{Nielsen2020}. The available data consists of 78 data points; Table~\ref{tab:data_IAD} lists the complete set of data points used.

% **********************************************************************
\subsubsection{\Gaia~DR2}
Unlike with \Hipp, the IAD for \Gaia~is not yet public. We thus utilise the reduced data consisting of five astrometric parameters: the right-ascension $\alpha_G$, declination $\delta_G$, the proper motion vectors $\mu_{\alpha,G}$ and $\mu_{\delta,G}$, and the parallax. Although the velocity in the data is recorded as a proper motion, in reality, in the presence of a companion, this velocity contains both the proper motion and the velocity induced by the movement of the photocenter around the barycenter of the system. The deviation of the velocity from the actual proper motion is often called the proper motion anomaly, which stems from the movement of the photocenter around the barycenter.

Due to \epsE's proximity and the long baseline between \Hipp~and \Gaia, we need to account for the 3D effects of different tangential planes where the PMs of each dataset are published. In other words, we account for the effect of the curvature when comparing the two coordinate systems. Correcting for this effect when propagating from the \Hipp's to \Gaia's epoch accounts for an error of $\sim$0.25 mas for the PM in the RA direction, which, although being below one sigma, is large enough to affect the model fit. We use the \texttt{SkyCoord}~tool in the \texttt{astropy}~Python package to propagate between epochs that accounts for the difference in epochs and position in spherical coordinates. This allows us to work combining velocities from both datasets. We follow a similar approach to \citet{Kervella2019}.

Furthermore, we correct for systematics from Gaia's rotating frame of reference according to the method described in \cite{Lindegren2018}. 

\subsubsection{\Gaia~eDR3}
{We use the \Gaia~eDR3 to do a different fit to the proper motion anomalies between the \Hipp~and \Gaia~proper motions. We use the calibrated proper motions from \citet{Brandt2021}. This catalog includes the proper motions from the \Hipp~epoch, from the \Gaia~eDR3 epoch, and the long term proper motion computed with the positions at the two epochs and the time difference. The difference between the \Hipp~or \Gaia~proper motion and the long baseline proper motion is the proper motion anomaly, which is caused by the presence of the planet. For this fit we do not use the \Hipp~IAD.}

%
% **********************************************************************
%
\subsection{Direct Imaging Data}\label{sec:di}
Building upon the work presented by \cite{Mawet2019}, we present additional Ms-band observations on the \epsE~system acquired in 2019 with Keck/NIRC2 and its vector vortex coronagraph \citep{Serabyn2017}.  The observations are summarized in \autoref{tab:obssummary}. Except for the night of the 2019-12-08, the new data was acquired using the newly installed infrared pyramid wavefront sensor \citep{Bond2020} instead of the facility Shack-Hartmann wavefront sensor. The alignment of the star behind the coronagraph is ensured at the 2 mas level by the quadrant analysis of coronagraphic images for tip-tilt sensing (QACITS; \citealt{Huby2015,Huby2017}). The telescope was used in pupil tracking mode to allow speckle subtraction with angular differential imaging \citep{Marois2008}.

\begin{figure}[t!]
   \begin{center}
   \begin{tabular}{cc} %% 
   \includegraphics[width=0.48\textwidth]{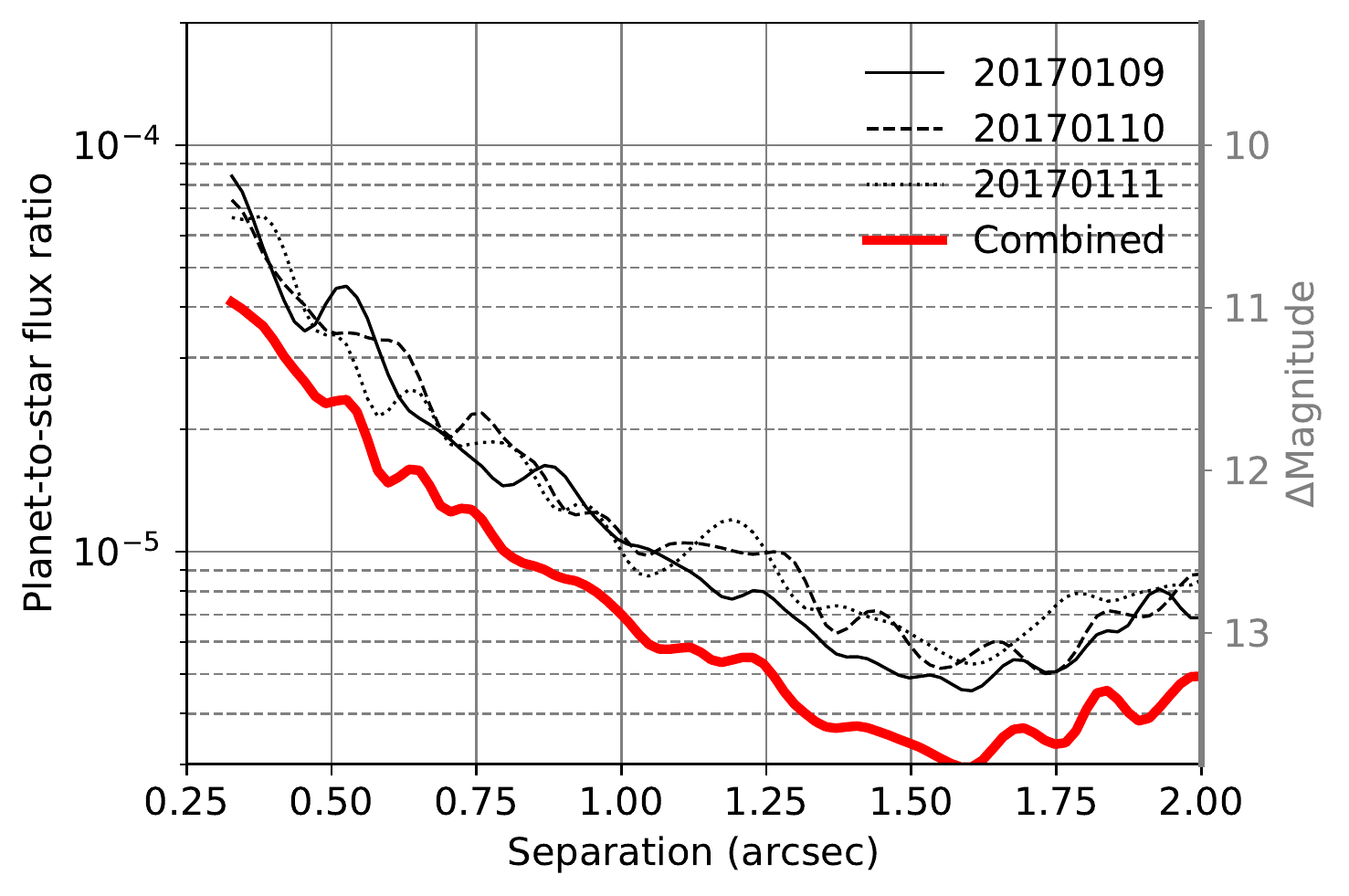} & 
   \includegraphics[width=0.48\textwidth]{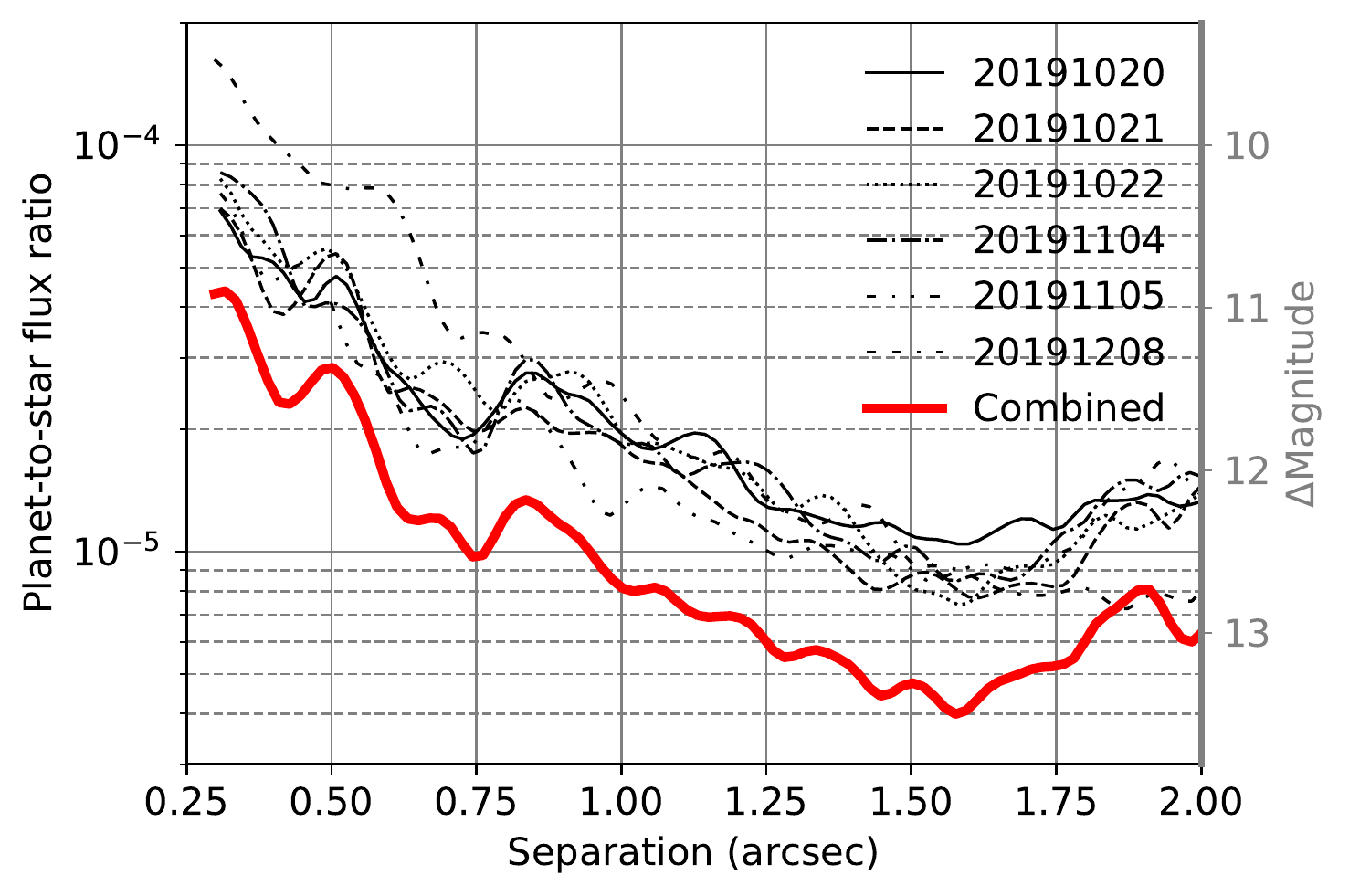} \\
   \end{tabular}
   \end{center}
   \caption{
   \label{fig:contrast}
   Planet sensitivity for each observations in 2017 (left) and 2019 (right). The planet sensitivity is expressed as the $5\sigma$ planet-to-star flux ratio. The December 8, 2019 epoch was not included in the combined sensitivity curve for 2019 as the planet would have moved by an amount comparable to the size of the point spread function.}
\end{figure}

The data was corrected for bad pixels, flat-fielded, sky-subtracted, and registered following the method detailed in \citet{Xuan18}. Similar to \citet{Mawet2019}, the stellar point spread function was subtracted using principal component analysis combined with a matched filter \citep[FMMF;][]{Ruffio2017} from the open-source Python package pyKLIP \citep{Wang2015}. After high-pass filtering the images with a Gaussian filter with a length scale of twice the PSF full width at half maximum (FWHM), the PSF FWHM is $\sim$10 pixels ($\sim$0.1$''$) 
in Ms band. Stellar speckles are subtracted using 30 principal components. For each science frame, the principal components are only calculated from images featuring an azimuthal displacement of the planet on the detector of more than 0.5 FWHM. The final sensitivity of each epoch is plotted in \autoref{fig:contrast} as a function of projected separation.
The nightly performance in 2019 was lower than in 2017 resulting in a similar combined sensitivity for each year despite the longer combined exposure time. A weighted mean is used to combine different epochs together. As a way to correct for any correlation between frames, the noise of the combined images is then normalized following the standard procedure of dividing by the standard deviation calculated in concentric annuli. With the exception of the {2019-12-08} observation, the epochs in 2017 and 2019 are close enough in time for the planet to not move significantly compared to the PSF size with a full width at half-maximum of $\sim$10 pixels.

\begin{deluxetable}{ccccccccc}
\tabletypesize{\scriptsize}
\tablecaption{Summary of observations of \epsE~with Keck/NIRC2 in Ms band. From left to right: UT date of the observation, total exposure time, exposure time per integration, number of coadds, number of exposures, total paralactic angle rotation for angular differential imaging, 5$\sigma$ sensitivity expressed as the planet-to-star flux ratio at 2 projected separation (0.4 and 1.0''), and finally the wavefront sensor (WFS) used (Shack-Hartmann or Pyramid). \label{tab:obssummary}}
\tablewidth{0pt}
\tablehead{
\colhead{Epoch} & \colhead{Tot. time } & \colhead{Exp. time} & \colhead{Coadds} & \colhead{\# of exp.} & \colhead{Angle rot.} & \colhead{5$\sigma$} & \colhead{5$\sigma$} & \colhead{WFS}\\
  & 
  \colhead{(s)} & 
  \colhead{(s)} & & &
  \colhead{(deg)} & 0.4'' & 1.0'' & 
}
\startdata
2017-01-09 & 6300 & 0.5 & 60 & 210 & 88.6 & 4.8e-05 & 1.1e-05 & SH\\
2017-01-10 & 7800 & 0.5 & 60 & 260 & 100.2 & 4.9e-05 & 1.1e-05 & SH\\
2017-01-11 & 4800 & 0.5 & 60 & 160 & 69.5 & 6.3e-05 & 1.0e-05 & SH\\
\hline
Combined 2017 & 5.25h &&&&& 3.3e-05 & 7.0e-06 & \\
\hline
\hline 
2019-10-20 & 5190 & 0.25 & 120 & 173 & 84.9 & 5.1e-05 & 2.0e-05& Pyramid\\
2019-10-21 & 5850 & 0.25 & 120 & 195 & 82.0 & 3.9e-05 & 1.8e-05& Pyramid\\
2019-10-22 & 7170 & 0.25 & 120 & 239 & 89.3 & 5.4e-05 & 2.0e-05& Pyramid\\
2019-11-04 & 6660 & 0.25 & 120 & 222 & 68.2 & 6.3e-05 & 1.8e-05& Pyramid\\
2019-11-05 & 6510 & 0.25 & 120 & 217 & 68.0 & 4.6e-05 & 1.3e-05& Pyramid\\
2019-12-08 & 6510 & 0.3 & 100 & 217 & 103.7 & 1.0e-04 & 2.4e-05& SH\\
\hline
Combined 2019 & 10.5h &&&&& 2.3e-05 & 7.9e-06 & \\
\enddata
\end{deluxetable}

%
% **********************************************************************
% **********************************************************************
% **********************************************************************
%
\section{Analysis}\label{sec:analysis}
% **********************************************************************

\subsection{Radial Velocity Model}\label{sec:rv_analysis}
\citet{Mawet2019} performed a thorough series of tests to evaluate the possibility that \epsE~b is an artifact of stellar activity, finding that the $\sim$7 yr orbital period is distinct from periods and harmonics of the periodicities in the S$_{HK}$ activity indicator timeseries. Our aim is not to recapitulate their analysis, but to update their orbital solution using the additional data obtained since the paper's publication, which spans approximately half of one orbital period of \epsE~b.  

\citet{Mawet2019} identified three peaks in a Lomb-Scargle periodogram of the RVs that rose above the 1\% eFAP: one at the putative planet period of 7.3 yr, one at 2.9 yr, and one at 11d. Applying \texttt{RVSearch} (Rosenthal et al. 2021) to the full dataset, we recover this structure of peaks. Like \citet{Mawet2019}, we also identify two major periods in the S$_{HK}$ timeseries for both HIRES and APF, which coincide with the 11 d and 2.9 yr periods in the RVs. We interpret these as the signatures of rotationally-modulated stellar activity and a long-term activity cycle, respectively. To investigate the effects of these signals on the physical parameters of planet b, we performed two separate RV orbit-fits using \texttt{RadVel} \citep{Fulton2018} to try different priors on the Gaussian Process (GP) timescale for modeling the stellar activity. In each of these fits, we assume a one-planet orbital solution, parameterized as $\sqrt{e_b}\cos{\omega_b}$,  $\sqrt{e_b}\sin{\omega_b}$, T$_{conj,b}$, P$_b$, K$_b$. We also included RV offset ($\gamma$) and white noise ($\sigma$) parameters for each instrument in the fit, treating the four Lick velocity datasets independently to account for instrumental upgrades as in \citet{Mawet2019}. Finally, we allowed an RV trend $\dot{\gamma}$.

In the first fit, we included a GP noise model to account for the impact of rotationally-modulated magnetic activity on the RVs \citep{Rajpaul2015}. We used a quasi-periodic kernel, following \citep{Mawet2019}. This kernel has hyperparameters $\eta_2$, the exponential decay timescale (analagous to the lifetime of active regions on the stellar disk), $\eta_3$, the stellar rotation period, and $\eta_4$, which controls the number of local maxima in the RVs per rotation period, and $\eta_1$, the amplitude of the GP mean function, which we treated as independent for each instrument dataset. Following \citet{Lopez-Morales2016}, we fixed $\eta_4$ to 0.5, which allows approximately two local maxima per rotation period. In this first fit, we allowed $\eta_2$ and $\eta_3$ to vary in the range (0, 100d). We calculated a Markov chain representation of the posterior using \texttt{emcee} \citep{Foreman-Mackey2013}, We visually inspected the chains to ensure appropriate burn-in and production periods. In total, the chain contained 450080 samples. The resulting orbital and nuisance parameters are given in Table \ref{tab:params}. Both the orbital parameters and the GP hyperparameters are well constrained; in particular, the marginalized posterior over the rotation period $\eta_3$ is Gaussian about the expected period of 11d. The data allow a trend, although the value is consistent with no trend at the 1$\sigma$ level, which allows us to conclude that there is no evidence in the current data for a RV trend. One non-intuitive feature of this fit is a preference for extremely small values ($10^{-6}$ m/s) of white noise jitter for the second Lick dataset. Even when we ran a fit requiring that all jitter values be at least 0.5 m/s, the posterior peaked at this lower bound. This may be evidence that much of the noise in this particular dataset is correlated, and therefore well-modeled by the GP noise model. It could also indicate that the reported observational uncertainties are overestimated for this dataset. Whatever the reason, there are fewer than 10 measurements that are affected by the value of this jitter parameter, and neither the orbital parameters nor the GP timescale parameters are affected by its particular value. 

To investigate the impact of the long-term activity cycle on the marginalized orbital parameter posteriors, we performed another fit identical to the one described above, except we required that $\eta_2$ and $\eta_3$ vary between 1yr and the $\sim30$ yr observation baseline. The marginalized posteriors for $\eta_2$ and $\eta_3$ were broad, with power across the entire allowed space, although the period parameter $\eta_3$ showed local maxima at both $\sim1100$ d and $\sim2000$ d. The marginalized 1 d posteriors for both of these parameters did not vary with those of any of the orbital parameters, allowing us to conclude that the long-term activity cycle, while somewhat present in the RVs, did not significantly affect the derived values of the orbital parameters. We therefore adopt the rotation-only GP fit described above.

\begin{figure}
    \centering
    \includegraphics{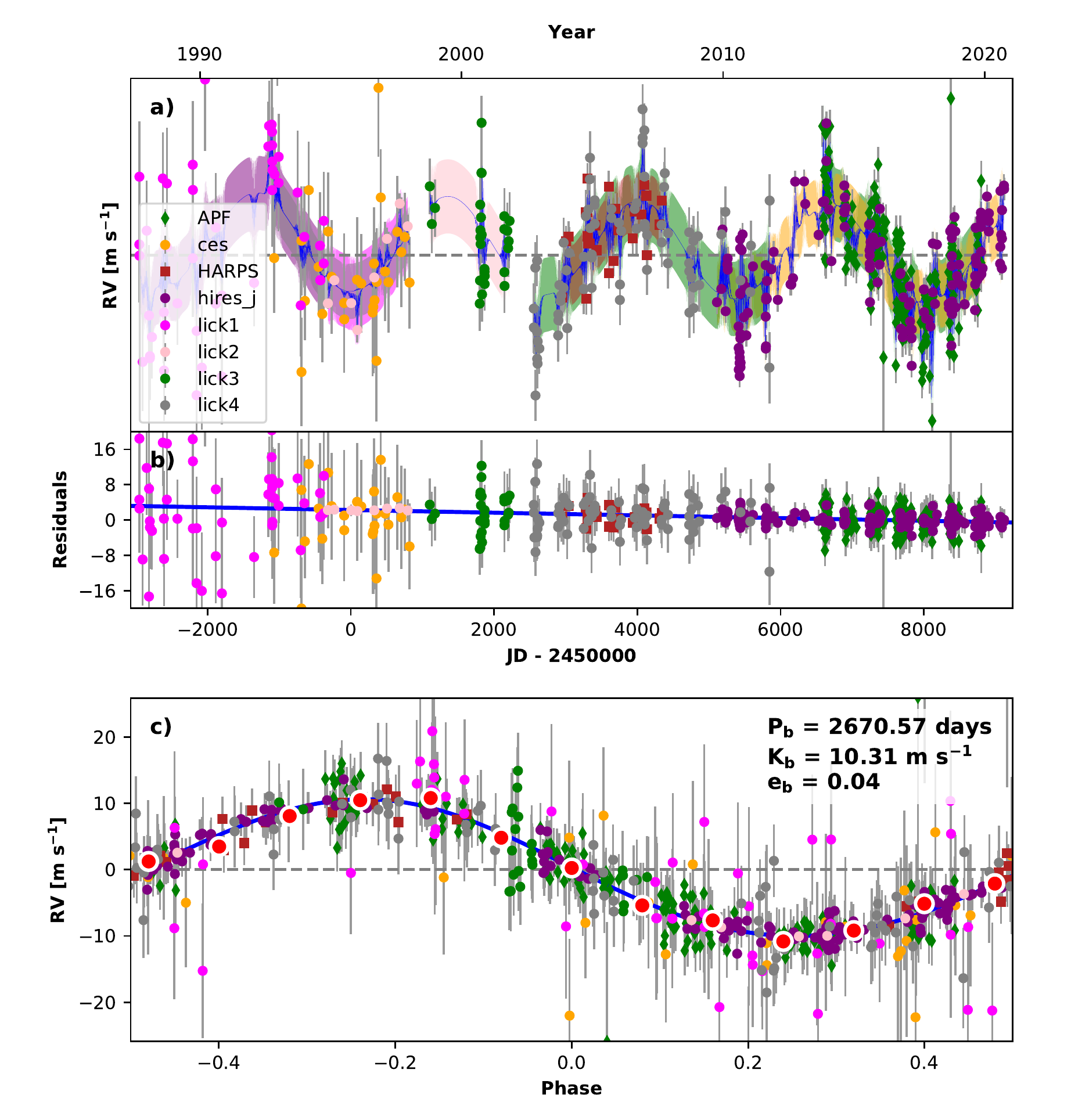}
    \caption{ (a) Time series of radial velocities from all data sets, (b) residuals to the RV fit, (c) phase-folded RV curve. The maximum probability one-planet model is overplotted (\textit{blue}), as well as the binned data (\textit{red dots}).}
    \label{fig:rv_multipanel}
\end{figure}

Our derived orbital parameters, {accounting} for the effect of rotationally-modulated stellar noise, are very similar to those of \citet{Mawet2019} (see their Table~3). We derive an orbital period of $2671^{+17}_{-23}$ days, a slightly reduced median semiamplitude of 10.3 m/s, and a low median eccentricity of 0.067. We show the series of RV measurements, the residuals to the fit and the phase folded RV curve in Fig.~\ref{fig:rv_multipanel}.

\begin{deluxetable}{lrrr}
\tablecaption{ RV Fit MCMC Posteriors }
\tablehead{
  \colhead{Parameter} & 
  \colhead{Credible Interval} & 
  \colhead{Maximum Likelihood} & 
  \colhead{Units}
}
\startdata
\sidehead{\bf{Modified MCMC Step Parameters}}
  $P_{b}$ & $2671^{+17}_{-23}$ & $2661$ & days \\
  $T\rm{conj}_{b}$ & $2460017^{+76}_{-32}$ & $2460023$ & JD \\
  $T\rm{peri}_{b}$ & $2460054^{+680}_{-690}$ & $2460235$ & JD \\
  $e_{b}$ & $0.055^{+0.067}_{-0.039}$ & $0.046$ &  \\
  $\omega_{b}$ & $57.3^{+80.2}_{-154.7}$ & $2.1$ & $^\circ$ \\
  $K_{b}$ & $10.34^{+0.95}_{-0.93}$ & $10.33$ & m s$^{-1}$ \\
\hline
\sidehead{\bf{Other Parameters}}
  $\gamma_{\rm lick4}$ & $-1.0^{+2.7}_{-2.6}$ & $-0.9$ & m s$^{-1}$ \\
  $\gamma_{\rm lick3}$ & $10.0\pm 4.9$ & $9.9$ & m s$^{-1}$ \\
  $\gamma_{\rm lick2}$ & $7.5^{+5.6}_{-5.7}$ & $7.6$ & m s$^{-1}$ \\
  $\gamma_{\rm lick1}$ & $4.4^{+6.2}_{-6.1}$ & $4.5$ & m s$^{-1}$ \\
  $\gamma_{\rm hires_j}$ & $2\pm 1$ & $2$ & m s$^{-1}$ \\
  $\gamma_{\rm harps}$ & $16442.5^{+3.2}_{-3.1}$ & $16442.4$ & m s$^{-1}$ \\
  $\gamma_{\rm ces}$ & $16446.6^{+5.7}_{-5.5}$ & $16446.6$ & m s$^{-1}$ \\
  $\gamma_{\rm apf}$ & $-0.9\pm 1.3$ & $-0.9$ & m s$^{-1}$ \\
  $\dot{\gamma}$ & $-0.00026^{+0.00063}_{-0.0006}$ & $-0.00026$ & m s$^{-1}$ d$^{-1}$ \\
  $\ddot{\gamma}$ & $\equiv0.0$ & $\equiv0.0$ & m s$^{-1}$ d$^{-2}$ \\
  $\sigma_{\rm lick4}$ & $5.17^{+1.1}_{-0.95}$ & $5$ & $\rm m\ s^{-1}$ \\
  $\sigma_{\rm lick3}$ & $5.6^{+2.1}_{-2.3}$ & $5.3$ & $\rm m\ s^{-1}$ \\
  $\sigma_{\rm lick2}$ & $2.3^{+3.6}_{-1.4}$ & $0.5$ & $\rm m\ s^{-1}$ \\
  $\sigma_{\rm lick1}$ & $7.2^{+3.5}_{-3.7}$ & $7.4$ & $\rm m\ s^{-1}$ \\
  $\sigma_{\rm hires_j}$ & $2.36^{+0.46}_{-0.41}$ & $2.26$ & $\rm m\ s^{-1}$ \\
  $\sigma_{\rm harps}$ & $4.8^{+2.2}_{-2.7}$ & $4.3$ & $\rm m\ s^{-1}$ \\
  $\sigma_{\rm ces}$ & $8.1^{+3.8}_{-4.5}$ & $7.3$ & $\rm m\ s^{-1}$ \\
  $\sigma_{\rm apf}$ & $3.64^{+0.62}_{-0.55}$ & $3.51$ & $\rm m\ s^{-1}$ \\
  $\eta_{1, \rm apf}$ & $7.82^{+0.97}_{-0.92}$ & $7.71$ & m s$^{-1}$ \\
  $\eta_{1, \rm hires_j}$ & $6.92^{+0.64}_{-0.59}$ & $6.78$ & m s$^{-1}$ \\
  $\eta_{1, \rm ces}$ & $7.5^{+4.1}_{-4.7}$ & $7.0$ & m s$^{-1}$ \\
  $\eta_{1, \rm harps}$ & $5.7^{+2.3}_{-3.0}$ & $5.3$ & m s$^{-1}$ \\
  $\eta_{1, \rm lick,{fischer}}$ & $8.7^{+1.3}_{-1.4}$ & $8.0$ &  \\
  $\eta_{2}$ & $37.6^{+6.4}_{-5.4}$ & $36.4$ & days \\
  $\eta_{3}$ & $11.68^{+0.14}_{-0.13}$ & $11.66$ & days \\
  $\eta_{4}$ & $\equiv0.5$ & $\equiv0.5$ &  \\
\enddata
\tablenotetext{}{ 436160 links saved}
\tablenotetext{}{
  Reference epoch for $\gamma$,$\dot{\gamma}$,$\ddot{\gamma}$: 2457454.642028 
}
\label{tab:params}
\end{deluxetable}

\subsection{Combining Direct Imaging, Astrometry, and Radial Velocity Data}\label{sec:using_rv}
We use the RV posterior distributions presented in Sec.~\ref{sec:rv_analysis} that we obtain from the \texttt{RadVel} orbit fit as the prior probabilities for the model fit to the astrometry and direct imaging data. The set of priors from the RV fit consists of: $a$, $e$, $\omega$, $\tau$, and $M~$sin~$i$. We use the same orbital parameter convention as in \citet{Blunt2020}. However, since the fit to the astrometry and direct imaging data will obtain a separate distribution for both the mass and inclination, we draw a set of correlated values for $i$ and $M$ from the $M~$sin~$i$ distribution. We assume a \textit{sine} distribution as a prior probability for $i$, which corresponds to an unconstrained prior for $i$.

Such a set of parameters presents a complex covariance that is practically impossible to reproduce if trying to represent these distributions in a parametric way. Instead, we choose to use a Kernel Density Estimator (KDE) to translate the posteriors to priors. A KDE smooths a probability density by convolving a Gaussian kernel with the discrete points of the MCMC chain. It allows one to preserve the covariance information contained in the original posteriors while allowing for a reliable reproduction of the shapes in the distributions.
We use the \texttt{scipy.stats} package implementation of a KDE, which allows for a customized value of the KDE bandwidth parameter. The choice of the bandwidth is critical to maintain the fidelity of the RV fit for the fit of the astrometry and direct imaging data. We explain the process to compute the optimal bandwidth for our data in Appendix~\ref{sec:kde}.

The combined astrometric and imaging model consists of thirteen parameters: the six orbital parameters ($a$, $e$, $i$, $\omega$, $\Omega$, $\tau$), the mass of the companion $M_b$, the total mass of the system $M_{tot}$, the parallax, the position of the barycenter of the system at an arbitrary epoch (we choose the \Hipp~epoch at 1992.25, $\alpha_{1992.25}$ and $\delta_{1992.25}$), and the proper motion vector of the barycenter $\mu = [\mu_{\alpha}, \mu_{\delta}]$.

\subsection{Astrometric Model}\label{sec:astrometry_model}
For the astrometric model, we use a similar approach to \cite{DeRosa2019} and \cite{Nielsen2020} while also fitting for the parallax and $M_{tot}$. This framework consists on deriving from the model a displacement of the photocenter from the barycenter induced by the presence of a companion. The orbital parameters and the relative mass are used to compute this relative motion of the photocenter about the barycenter. Then, to compare to the astrometry data, we obtain absolute astrometry quantities propagating from a reference RA/Dec position with the proper motion of the barycenter, and adding the relative displacement. For this reason we add to our model the reference position $\alpha_{1992.25}$ and $\delta_{1992.25}$, and the proper motion of the barycenter $\mu$.
We assume that the brightness of the planet is negligible compared to the host star (below 10 ppb reflected light in the visible); consequently making the photocenter and the star positions coincide.

We compute the goodness of fit to the astrometric data by deriving two distinct terms for both the \Hipp~and \Gaia~data: $\chi^2_{astrom} = \chi^2_{H} + \chi^2_{G}$. 
For the \Hipp~IAD, we calculate the expected position of the barycenter at all IAD epochs ($\alpha_H$ and $\delta_H$) by propagating $\alpha_{1992.25}$ and $\delta_{1992.25}$ with $\mu$. The displacement of the photocenter with respect to the barycenter ($\Delta\alpha_H$ and $\Delta\delta_H$) is computed using the expected orbit from the model orbital parameters, and the relative mass. We compare this values to the corresponding $\alpha_{IAD}$ and $\delta_{IAD}$ that we calculate from the 1D scans from IAD. We compute $\chi^2_{H}$ in the same way as in \cite{Nielsen2020}.

Similarly for the \Gaia~data, we propagate the reference position of the barycenter, $\alpha_{1992.25}$ and $\delta_{1992.25}$, to the \Gaia~epoch: $\alpha_G$ and $\delta_G$, and we compute the displacement of the photocenter with respect to the barycenter ($\Delta\alpha_G$ and $\Delta\delta_G$) with the orbital parameters and the relative mass. We compare this derived quantity to the {\Gaia~DR2} positional data, i.e. the RA/Dec from the {\Gaia~DR2} catalog $\alpha_{G,GDR2}$ and $\delta_{G,GDR2}$. We can fit to the {\Gaia~DR2} proper motion data by adding an instantaneous proper motion disturbance to the reference proper motion calculated by deriving an average linear velocity of the star due to its orbital motion over a few epochs around 2015.5. %This proper motion disturbance is noted $\Delta\mu_{G,\alpha}, \Delta\mu_{G,\delta}$. 
The $\chi^2_G$ associated with the \Gaia~data is then calculated in the same way as described in \citet{DeRosa2019}. 

{The analysis of the \Hipp-\Gaia~eDR3 acceleration data is done in the way described in \citet{Kervella2019}. }

\subsection{Direct Imaging Model}
The part of the likelihood function associated with the direct imaging data described in Sec.~\ref{sec:di} is computed based on the method described in \citet{Mawet2019}. The logarithm of the direct imaging likelihood ($\mathcal{P}$) \citep{Ruffio2018} for a single epoch can be written as:

\begin{equation}
    \log \mathcal{P}(d|I,x) = -\frac{1}{2\sigma^2_x}(I^2-2 I \tilde{I}_x)
\end{equation}

Where $I$ is the planet flux corresponding to the planet mass in the orbital model, $\tilde{I}_x$ is the estimated flux from the data at the location $x$, and $\sigma_x$ is the uncertainty of this estimate. Individual epochs are not combined in the main analysis of this work. The direct imaging epochs are assumed to be independent such that their log-likelihoods are simply added together. While this assumption is not perfect, it should not significantly affect the upper limit as the framework marginalizes over the spatial direction which will factor in any brighter speckles. To compute $I$ from the planet mass, we use the COND evolutionary model \citep{Baraffe2003} and we adopt an upper bound age of 800 Myr for the system \citep{Janson2015}. To compare with this age, we also run model fits with an age of 400 Myr, corresponding to the lower bound. $\tilde{I}_x$ is the flux measured in the image where the planet is predicted to be based on a given set of orbital parameters.

% **********************************************************************
\subsection{MCMC Results}\label{sec:mcmc}

The fit to the astrometry and direct imaging data using the \texttt{RadVel} posteriors as priors is performed using the \orbitize\footnote{https://github.com/sblunt/orbitize} package \citep{Blunt2020}. We implement a Markov-chain Monte Carlo (MCMC) \citep{Foreman-Mackey2013} analysis to fit for the six orbital parameters and the mass of planet b. {MCMC has been extensively used to do orbit fitting, e.g. first by \citet{Hou2012} and more recently by \citet{Blunt2019},~\citet{Nowak2020},~\citet{Hinkley2021} and \citet{Wang2021}. } 
As discussed in Sec.~\ref{sec:using_rv} we also fit for the mass of \epsE, as well as for the astrometric parameters $\alpha$ and $\delta$, and $\mu_{\alpha}$ and $\mu_{\delta}$. Therefore we fit for thirteen parameters; six orbital parameters, the parallax, the photocenter position vector and the proper motion vector, and the masses of the star and planet. 

The knowledge of \epsE's mass is accounted with a Gaussian prior $0.82\pm0.02~M_\odot$ \citep{Baines2012}. The priors for the astrometric parameters are set to uniform distributions centered at the \Hipp~values with broad ranges of $\pm20\sigma$ to allow for deviations of the data from the nominal proper motion and photocenter position. The RV fits do not provide any constraints on the longitude of the ascending node, $\Omega$, for which we set a uniform prior distribution. As discussed in Sec.~\ref{sec:using_rv}, although the RV fits do not constrain the inclination, $i$, we assume a \textit{sine} distribution to draw correlated values for the mass, $M_b$, and the inclination. The fit is performed with 550 walkers, and 18000 steps per walker.

The MCMC fits converge to yield the orbital parameters and mass posteriors shown in Table~\ref{tab:post}. Two fits are presented for both ends of the current bounds on the system's age, i.e., 400 and 800 Myr. A corner plot showing some of these and their correlations, for the 800 Myr fit, is shown in Fig.~\ref{fig:corner_plot}. A complete corner plot is shown in Appendix~\ref{sec:app_cornerplot}. 

In order to assess the constraining power of the direct imaging data, given that our data consists of nondetections, we performed an MCMC run to only fit for the astrometry data. Fig.~\ref{fig:mass_post} show the posterior distribution compared to the prior distribution, i.e. the posteriors from the RV fit. 

The use of the astrometry data results in a significantly improved constraining of the inclination of the planet with respect to the results presented in \cite{Mawet2019}. 
Starting with a \textit{sine} distribution as the prior, which is centered at 90$^\circ$, the walkers converge onto an inclination of $i$ = $78.81_{-22.41}^{\circ+29.34}$. The addition of the astrometry data also yields a different model for the mass of the companion; by using the RV and direct imaging data we obtain a distribution of $M_b$ = $0.73_{-0.13}^{+0.34}$~M$_{Jup}$ (800 Myr), by adding the astrometry,  $M_b$ = $0.66_{-0.09}^{+0.12}$~M$_{Jup}$ (800 Myr). The astrometry data from \Hipp~and \Gaia~seem to favor a lower mass planet and more edge-on inclinations. {We compute the most probable perturbation semi-major axis: $\alpha_A=0.89$~mas, which is a factor of $\sim$2 away from the result reported by \citet{Benedict2006}, i.e. 1.88~mas. More details on the perturbed orbit can be found in Appendix~\ref{sec:app_perturbed}.}

{We add the fit to the \Hipp-\Gaia~eDR3 acceleration in the last column of Table~\ref{tab:post}. The results of this fit are largely consistent with the \Hipp~IAD-\Gaia~DR2 fits. Two main differences can be appreciated: (1) the inclination median is $\sim$90$^\circ$; however, the posterior probabilities converge to a similar inclination with respect to our other fits, i.e. $\sim$75$^\circ$, and to its supplementary angle, i.e. $\sim$105$^\circ$. This is probably because the rotation information available from the IAD is not accessible when using the proper motion anomalies. (2) The upper mass bound (see Fig.~\ref{fig:mass_post}) is slightly lower, which brings the mass to a lower mass solution.}

This result is an order of magnitude better than previously published \citep{Mawet2019}. The {argument of periapsis}, $\omega$, reported in \citet{Mawet2019} is the stellar $\omega$, which is the reason it is $\sim$180$^\circ$ off with respect to the result presented here.

%disk\citep{Janson2015}
\begin{deluxetable}{ l c c c c c}
\tabletypesize{\scriptsize}
\tablecaption{ MCMC posteriors for different sets of data and assumptions for the age of the system.  \label{tab:post}}
\tablewidth{0pt}
\tablehead{
  \colhead{}& 
  \colhead{RV and Direct Imaging} & 
  \multicolumn{2}{c}{RV, Astrometry, }  &  
  \colhead{RV and Astrometry} &
  \colhead{RV, Astrometry (eDR3 acc.),} \\
  {}& {}& \multicolumn{2}{c}{and Direct Imaging} &{}&{and Direct Imaging}\\
  \colhead{} & 
  \colhead{800 Myr} & 
  \colhead{400 Myr} &
  \colhead{800 Myr} & 
  \colhead{\textit{No age assumption}} &
  \colhead{800 Myr}
}
\startdata
$M_b$ ($M_{Jup}$)&  $0.74_{-0.15}^{+0.37}$
  &  $0.66_{-0.10}^{+0.11}$
  & $0.66_{-0.09}^{+0.12}$
  & $0.65_{-0.09}^{+0.10}$
  & {$0.64_{-0.09}^{+0.09}$} \\ 
  $a_b$ (AU) &$3.52_{-0.04}^{+0.04}$
  & $3.53_{-0.03}^{+0.04}$ 
  & $3.53_{-0.04}^{+0.03}$
  & $3.53_{-0.04}^{+0.04}$
  & {$3.52_{-0.04}^{+0.04}$} \\ 
  $e$ &  $0.07_{-0.05}^{+0.07}$
  & $0.07_{-0.05}^{+0.07}$
  & $0.07_{-0.05}^{+0.07}$
  & $0.07_{-0.05}^{+0.07}$
  & {$0.07_{-0.05}^{+0.07}$} \\ 
  $\omega$ ($^\circ$) &  $-29.01_{-111.95}^{+107.96}$
  & $-28.37_{-99.58}^{+75.90}$
  & $-19.15_{-94.80}^{+88.27}$
  & $-19.54_{-87.06}^{+97.09}$
  & { $-29.84_{-115.85}^{+104.59}$} \\ 
  $\Omega$ ($^\circ$) &  $181.27_{-125.02}^{+124.60}$
  & $195.06_{-72.21}^{+127.20}$
  & $198.18_{-63.14}^{+127.29}$
  & $195.08_{-73.72}^{+122.64}$
  & {$190.06_{-151.74}^{+108.72}$} \\ 
  $i$ ($^\circ$) &  $89.15_{-43.16}^{+45.03}$
  & $75.77_{-19.94}^{+29.47}$
  & $78.81_{-22.41}^{+29.34}$
  & $80.95_{-21.39}^{+27.55}$
  & {$89.70_{-25.08}^{+25.49}$} \\ 
  $\tau_{peri}$ &  $0.52_{-0.35}^{+0.33}$
  & $0.42_{-0.29}^{+0.34}$
  & $0.35_{-0.24}^{+0.33}$
  & $0.37_{-0.25}^{+0.32}$
  & {$0.52_{-0.36}^{+0.34}$} \\ \hline
  $\Delta\alpha_{H}$ (mas) &  %%
  & $-0.33_{-0.41}^{+0.54}$
  & $-0.25_{-0.47}^{+0.53}$
  & $-0.20_{-0.48}^{+0.51}$
  & \\ 
  $\Delta\delta_{H}$ (mas)&  %%
  & $0.27_{-0.46}^{+0.32}$
  & $0.30_{-0.47}^{+0.32}$
  & $0.25_{-0.49}^{+0.32}$
  & \\ 
  $PM^{\alpha}_{H}$ &  %%
  & $-975.20_{-0.02}^{+0.02}$ 
  & $-975.20_{-0.02}^{+0.02}$
  & $-975.20_{-0.02}^{+0.02}$
  & \\ 
  $PM^{\delta}_{H}$ &  %%
  & $19.95_{-0.02}^{+0.02}$ 
  & $19.95_{-0.02}^{+0.02}$ 
  & $19.95_{-0.02}^{+0.02}$
  & \\ % 
  \enddata
\end{deluxetable}

\begin{figure}[t!]
   \begin{center}
   \begin{tabular}{c} %% 
   \includegraphics[height=12.0cm]{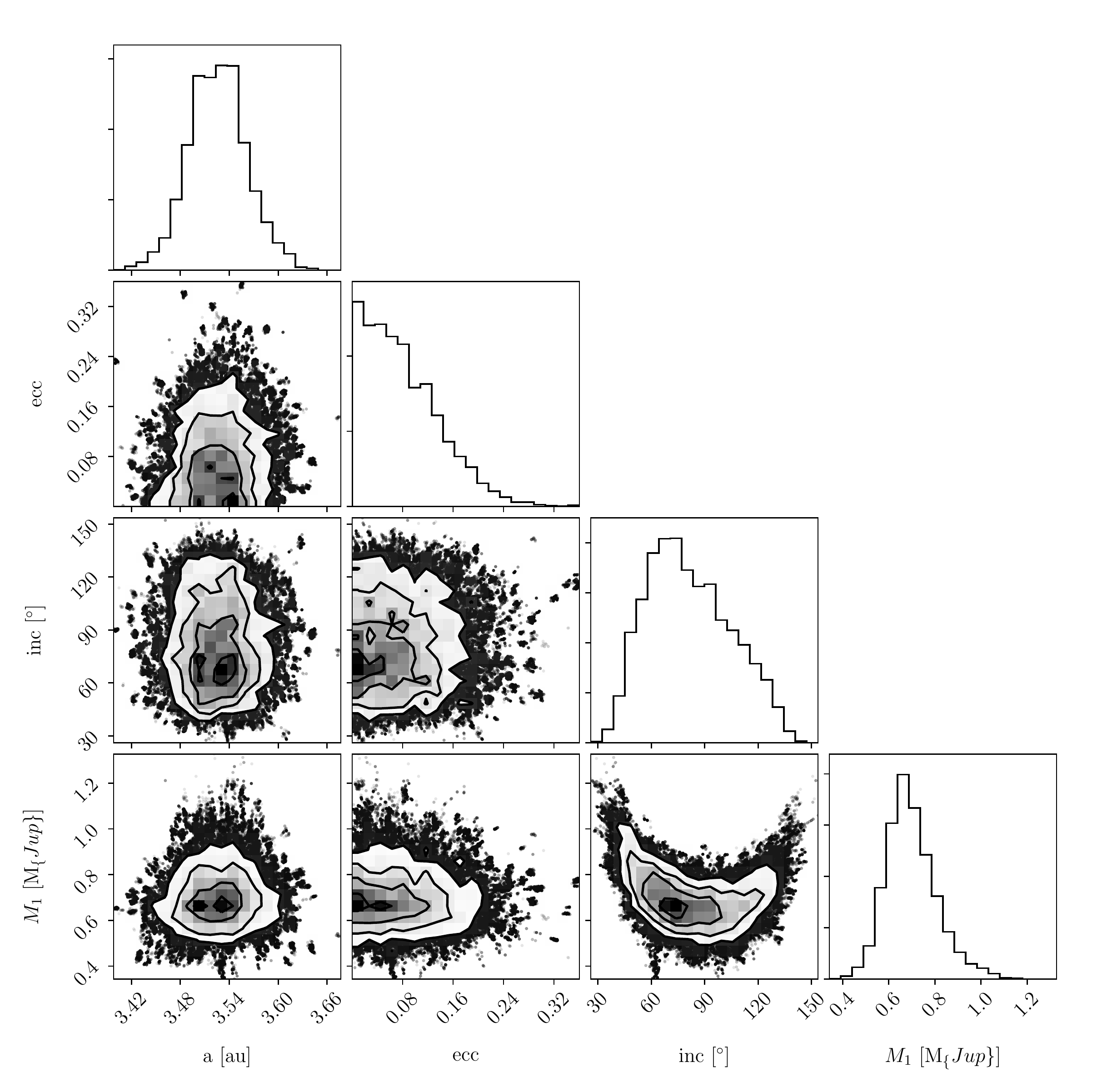}
   \end{tabular}
   \end{center}
   \caption{
   \label{fig:corner_plot}
   Corner plot of the posterior distributions and their correlation. These are the posteriors for the model fit to the RV, astrometry, and direct imaging data assuming an age of 800 Myr. We make use of \texttt{corner.py} \citep{Foreman-Mackey2016} to produce corner plots.} 
\end{figure}

% Figure: Posteriors vs Priors draw (= RV fit posteriors) 
\begin{figure}[t!]
   \begin{center}
   \begin{tabular}{c} %% 
   \includegraphics[height=7.5cm]{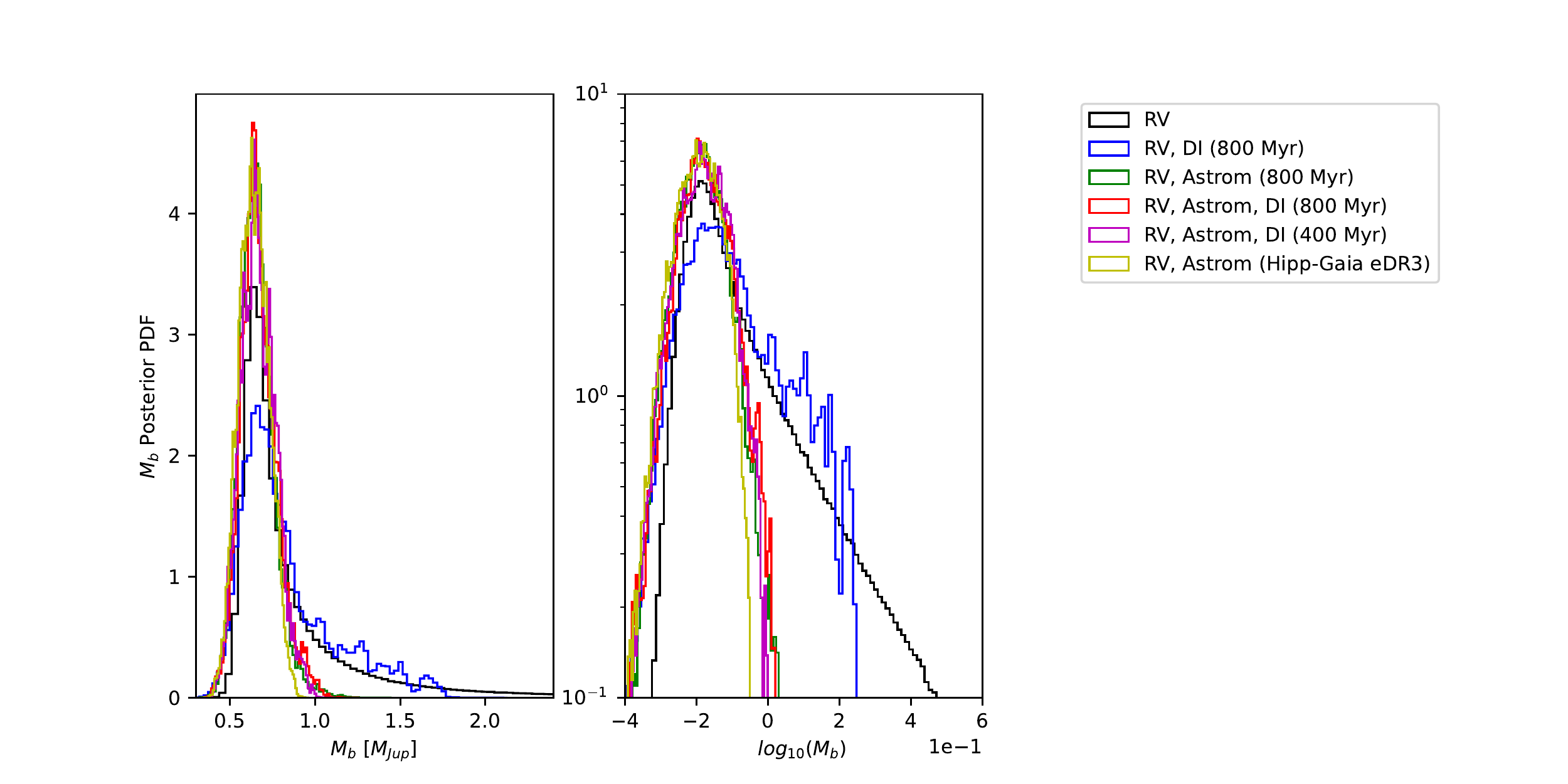}
   \end{tabular}
   \end{center}
   \caption{
   \label{fig:mass_post}
   Mass posteriors PDF for different fits in linear (\textit{left}) and logarithmic (\textit{right}) scale. The astrometry data has a bigger constraining power on both ends of the mass bounds with respect to the direct imaging data available. In all fits except the RV only fit (\textit{black}) we have utilized the KDE to include the RV posteriors as priors to the fit; an artifact thus appears at the extreme of lower bound  in which the distribution slightly separates from the prior. A detailed explanation of this can be found in Appendix~\ref{sec:kde}. However, closer to the median, the effect of adding the astrometry, for which a lower mass is allowed, is real, as it can be appreciated in the difference between the distributions with and without astrometry.  } 
\end{figure}

% Figure: Contour plots over Keck Data
\begin{figure}[t!]
   \begin{center}
   \begin{tabular}{c} %% 
   \includegraphics[height=7.5cm]{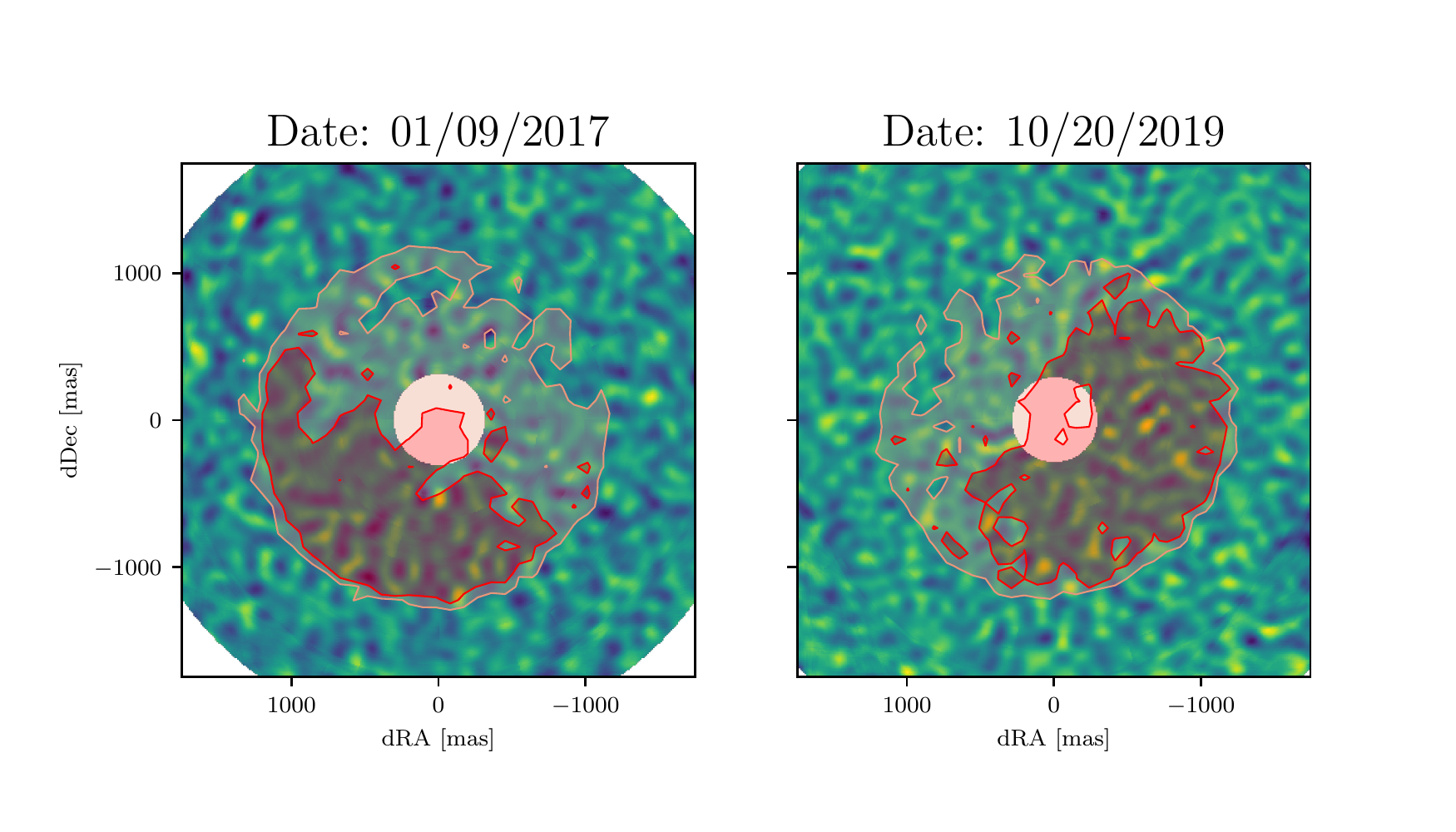}
   \end{tabular}
   \end{center}
   \caption{
   \label{fig:contour}
   Keck/NIRC2 reduced data for two of the nine observing epochs (see Table~\ref{tab:obssummary}), the 1- and 2-$\sigma$ contours of the posteriors of the position are overplotted. The posteriors are for the model fit to the RV, astrometry and direct imaging data assuming an age of 800 Myr. The addition of the astrometry and direct imaging data breaks the degeneracy on the inclination, and the position of the companion is better constrained.} 
\end{figure}

%
% **********************************************************************

\section{Discussion}\label{sec:discussion}
\subsection{Debris Disk}\label{sec:debrisDisk}
The interaction between the debris disk and planet in the \epsE~ system was thoroughly discussed in \citet{Mawet2019}. Here we build on this analysis with our updated orbital constraints. \epsE's debris disk is currently known to be composed of three rings \citep{Backman2008}: a main ring from 35-90 AU, an inner belt at $\sim$3 AU, an intermediate belt at $\sim$20 AU. Planet b sits between the two inner belts. Extensive characterization of the disk has been carried out over the years; see \citet{Backman2008,MacGregor2015,Booth2017}. In particular, \citet{Booth2017} analysis concluded the presence of a gap in emission in the circumstellar disk between $\sim$20 and $\sim$60 AU, which would indicate the presence of one or more companions in this range of separations. The inclination of the main ring was constrained to $34^\circ\pm2$ \citep{Booth2017}.

The orbital parameters and mass of planet b can set constraints on the edges of the inner belts \citep{Wisdom1980}. We follow the same analysis as \citet{Mawet2019} given that the eccentricity of the planet is expected to be well {below} 0.3, in which case the chaotic zone structure carved by the planet is independent of the eccentricity \citep{Quillen2006}. With the results of our MCMC fit (see Table~\ref{tab:post}), in particular the semi-major axis and relative mass, we expect there to be no particles from 2.97 to 4.29 AU. The edges are slightly moved outwards with respect to \citet{Mawet2019} since the semi-major axis posterior of the planet is now larger, it was then reported no particles from ~2.7 and~4.3 AU. The mass posterior is lower, which reduces the width of the chaotic zone by $\sim$4\%.

It was concluded in \citet{Mawet2019} that the inner belt and outer ring were most likely to be self-stirred, i.e. collisions between disk particles are driven by particle-to-particle gravitational interactions, as opposed to being stirred by the presence of the planet (see Sec.~5.3.2. and Fig.~15 in \citet{Mawet2019}). The timescale for the planet to stir and shape the main ring is $\sim$1 Gyr, which, combined with the distance between the two, makes their dynamical coupling probably not significant. As for the intermediate belt, the timescales of self-stirring and planet-stirring were computed to be similar in the range of separations of the belt. The new orbital parameters from our MCMC fit (see Table~\ref{tab:post}), namely a lower mass for the planet and a slightly higher semi-major axis, contribute to larger planet-stirring timescales: an 18\% increase in the planet-stirring timescale estimation with respect to \citet{Mawet2019}. However, this does not rule out the contribution of planet induced stirring process for the intermediate belt.
% Therefore, the resulting mutual inclination is 

The inclination of \epsE~b is constrained to $75.77_{-21.32}^{\circ+29.92}$ thanks to the inclusion of the astrometry data (see Sec.~\ref{sec:mcmc}). This result indicates that the planet orbit is likely inclined with respect to the main ring, for which $i = 34^\circ\pm2^\circ$, which is $\sim$2$\sigma$ away from the most probable inclination. 
The origin of such a mutual inclination is unknown but could possibly be due to the dynamical effects of a third body. 

A mutual inclination could be causing a warping on the main ring. Indeed, the vertical warp in the $\beta$ Pictoris inner disk is believed to have been produced by its mutual inclination with $\beta$ Pictoris~b \citep[e.g.,][]{Dawson2011}. Using a similar analysis as presented in \citet{Dawson2011} based on secular interactions, we find that in the case of \epsE~the minimum mass of planet at 3 AU to excite the inclination of dust particles in the ring at 70 AU after 800 Myr is of $\sim$0.5 $M_{Jup}$. This indicates that planet b could be in the regime of starting to drive a warp in the main ring if it is indeed misaligned with the disk plane. A coplanar solution is still allowed by the data since an inclination of 32$^\circ$ is $\sim$1$\sigma$ away from the most probable inclination of the planet. {It is worth noting that \citet{Benedict2006} yielded a solution for the inclination of the companion of $30.1^\circ\pm3.8^\circ$.}

\subsection{Gaia's Future Sensitivity}\label{sec:gdr3}
\citet{DeRosa2020} recently published the prospects for constraining mass of 51 Eridani b with \Gaia's final data release. They simulated sets of \Gaia~data with different astrometric error estimates, and found that the detection was possible only with optimistic mass and astrometric uncertainties. We performed a similar analysis computing the astrometric signal of \epsE~b and comparing it to the sensitivity results of Fig.~11 on \citet{DeRosa2020}. We expect \epsE~to have a similar astrometric error due to its brightness since the brightness of the star sets the uncertainty in the scans. We get an estimate for this at \citet{Lindegren2018}: $\sim$50 $\mu as$. 

We find that the amplitude of the astrometric signal for \epsE~b is an order of magnitude stronger than that of 51 Eridani b. Indeed, the shorter distance to the system and the period the planet both favorable factors for a stronger astrometric signal. For the nominal \Gaia~mission span of 5 years and for an astrometric uncertainty in the scans of 50~$\mu a$, \epsE~b is detectable at $\sim$1~$M_{Jup}$, which falls on the higher end of our mass posterior probability. However, for the extended mission span of 8 years, for which 51 Eridani is only detectable for a high mass estimate, \epsE~b is readily detectable even at $\sim$0.5~$M_{Jup}$, which is on the lower than the median mass of the posterior probability presented in this paper. 

The final data release of Gaia's mission is a particularly exciting prospect for the exoplanet science field. Gaia final release will probably have access to constraining the dynamic mass of \epsE~b. However, as the work presented in this paper aims at showing, it is by using this data combined with other observations that the best science is attainable.  

\subsection{Advantages of Combining Different Methods}
The results presented in this paper are another example of the power of combining different methods to constrain a system's characteristics. By adding the astrometry data, we have identified new constraints for the inclination and longitude of the ascending node, both of which are inaccessible to an RV orbit fit. The direct imaging data, although it being a nondetection, sets upper limits on the mass. The distribution of most likely planet positions shown in Fig.~\ref{fig:contour} and how it prefers certain areas and fluxes is no coincidence; the MCMC walkers converge easier where the estimated flux from the coronagraph images is higher.

Although the constraints on the position shown in Fig.~\ref{fig:contour} are far from ideal, direct imaging planet hunters will take advantage of any position knowledge however small it may be. Indeed, data reduction techniques in direct imaging greatly benefit from a prior knowledge of the position of the object. For instance, in principal component analysis (PCA)\citep{Soummer2012} based methods, a great deal of speckle subtraction power is gained by treating the data by patches; knowing where the planet is more likely to be reduces the computing time and allows the algorithm to focus on a constrained area of interest. 

The synergies between RV, astrometry, and direct imaging data are currently being explored, and more work is being done in this direction \citep{Nowak_etal2020}.

% **********************************************************************
%
\subsection{Prospects for a Direct Detection with JWST}

\begin{figure}[t!]
   \begin{center}
   \begin{tabular}{c} %% 
   \includegraphics[height=7.2cm]{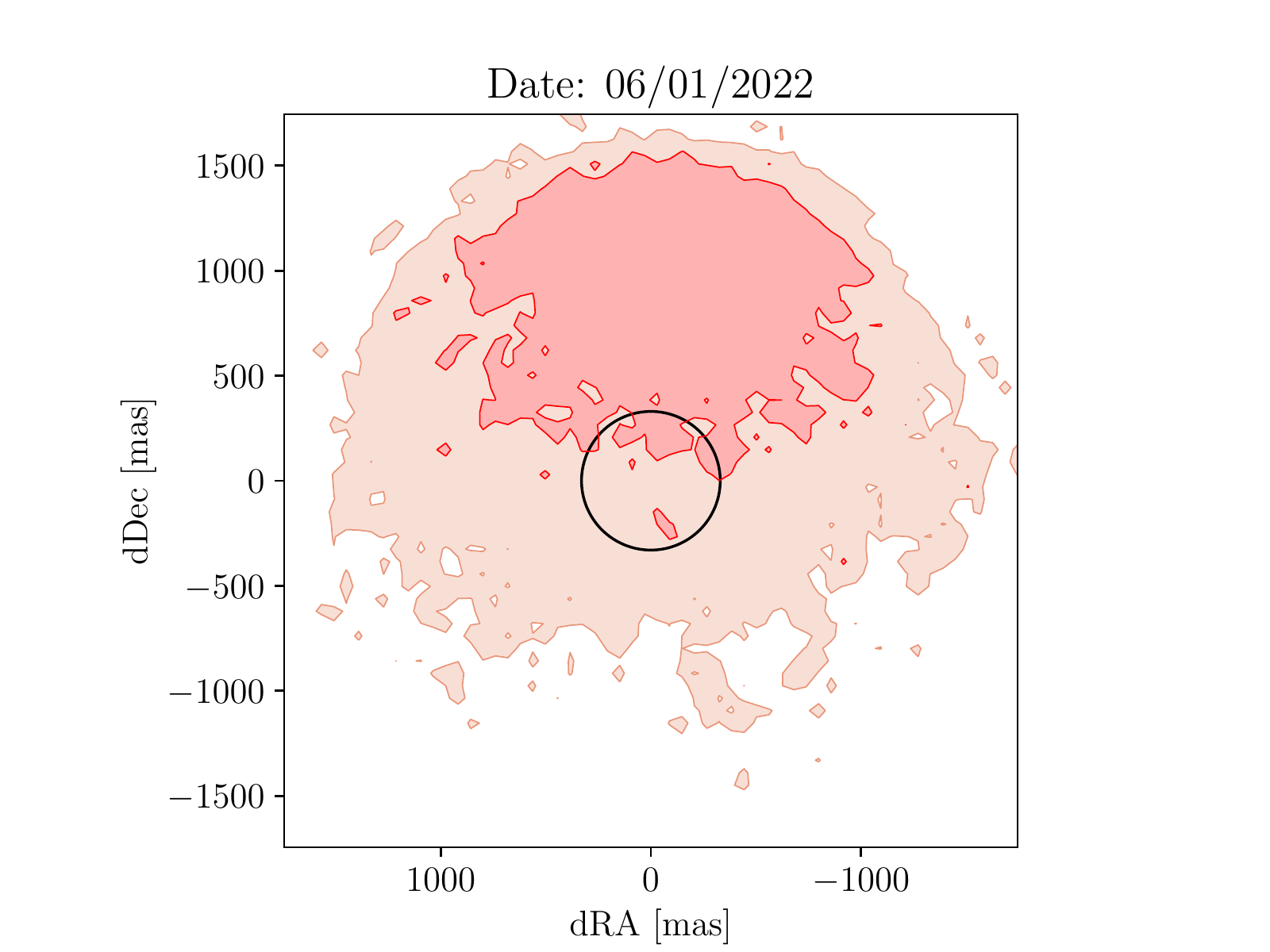}
   \end{tabular}
   \end{center}
   \caption{
   \label{fig:jwstcontour}
   1- and 2-$\sigma$ contours of the planet's position at a possible JWST epoch. This information can be useful for observers; for instance, when using the 4QPM coronagraph, the user will want to avoid the \textit{gaps} falling on the most probable position of planet b. The black circle indicates the inner working angle of MIRI's F1065 mode \citet{Boccaletti2015}. The posteriors used for this contours are taken from the fit to the RV, astrometry, and direct imaging data, assuming an 800 Myr age. } 
\end{figure}

The James Webb Space Telescope (JWST) will provide the community with unprecedented capabilities to do infrared exoplanet science. In this section we discuss the prospect for a detection of \epsE~b with JWST's NIRSPEC and MIRI instruments. In Fig.~\ref{fig:jwstcontour} we show the probability contour for the position of the planet at an epoch in JWST's Cycle 1. 

% Figure: Countour plots JWST

% **********************************************************************
\subsubsection{MIRI}
The Mid-Infrared Instrument (MIRI) aboard JWST is equipped with a coronagraph, and is expected to reach $10^{-4}$ levels of raw contrast at $\lambda =$ 10 $\mu m$ (mode F1065C) with conventional star subtraction \citep{Boccaletti2015}. We performed a more sophisticated data reduction as was done for MIRI in \citet{Beichman2020} to get a more accurate representation of the expected performance of MIRI on \epsE. We use the \texttt{IDL} library \texttt{MIRImSIM}\footnote{https://jwst.fr/wp/?p=30}, and the wavefront error drift predictions presented in \citet{Perrin2018}. We make use of as much diversity as possible from the jitter of the telescope to perform a reference star subtraction based on principal component analysis (PCA, \citet{Soummer2012}). A more detailed description of the data processing can be found in \citet{Beichman2020}. 

In Fig.~\ref{fig:nirspec_miri} we show some simulation results for MIRI observations. We find that, even for a perfect pointing accuracy of the telescope, MIRI would require $\sim$75 hours to reach SNR$>$5.% detection. 

% **********************************************************************
% **********************************************************************
\subsubsection{NIRSpec}
A new avenue to imaging \epsE~b is by using the moderate spectral resolution integral field spectroscopy mode of NIRSPEC (G395H/F290LP, with its 3x3’’ field of view). Atmospheric models predict a large excess emission around $4.5\,\mu$m from the atmosphere of exoplanets such as \epsE~b \citep{Marley2018sonora}. Ground-based medium resolution spectrographs ($R\sim4000$) like OSIRIS and SINFONI have made clear detections of exoplanets as close as $0.4''$ from their star detector water and carbon monoxyde \citep{Konopacky2013,Barman2015,Hoeijmakers2018,Wilcomb2020}. The advantage of a higher spectral resolution is the possibility to subtract the starlight continuum with a high-pass filter and then use cross-correlation techniques to detect the molecular spectral signature of the planet. Like NIRSpec, these instruments were not designed for exoplanet detection, but the increased resolution can overcome a lack of a coronagraph and achieve comparable, if not better, detections of imaged exoplanets. Furthermore, the fact that these are spectroscopic detections opens up rich capabilities in atmospheric characterization that are simply not possible through imaging alone. \citet{Houlle2021} demonstrated the power of high spectral resolution integral spectroscopy in the context of HARMONI, a first light instrument to the extremely large telescope, showing that it could detect planets 10 times fainter than angular differential imaging. NIRSpec is expected to excel at this technique thanks to the stability of a space observatory and the absence of variable telluric lines, which can be the source of spurious detection of molecules as discussed in \citet{PetitditdelaRoche2018}.

\begin{figure}[t!]
   \begin{center}
   \begin{tabular}{c} %% 
   \includegraphics[height=9.0cm]{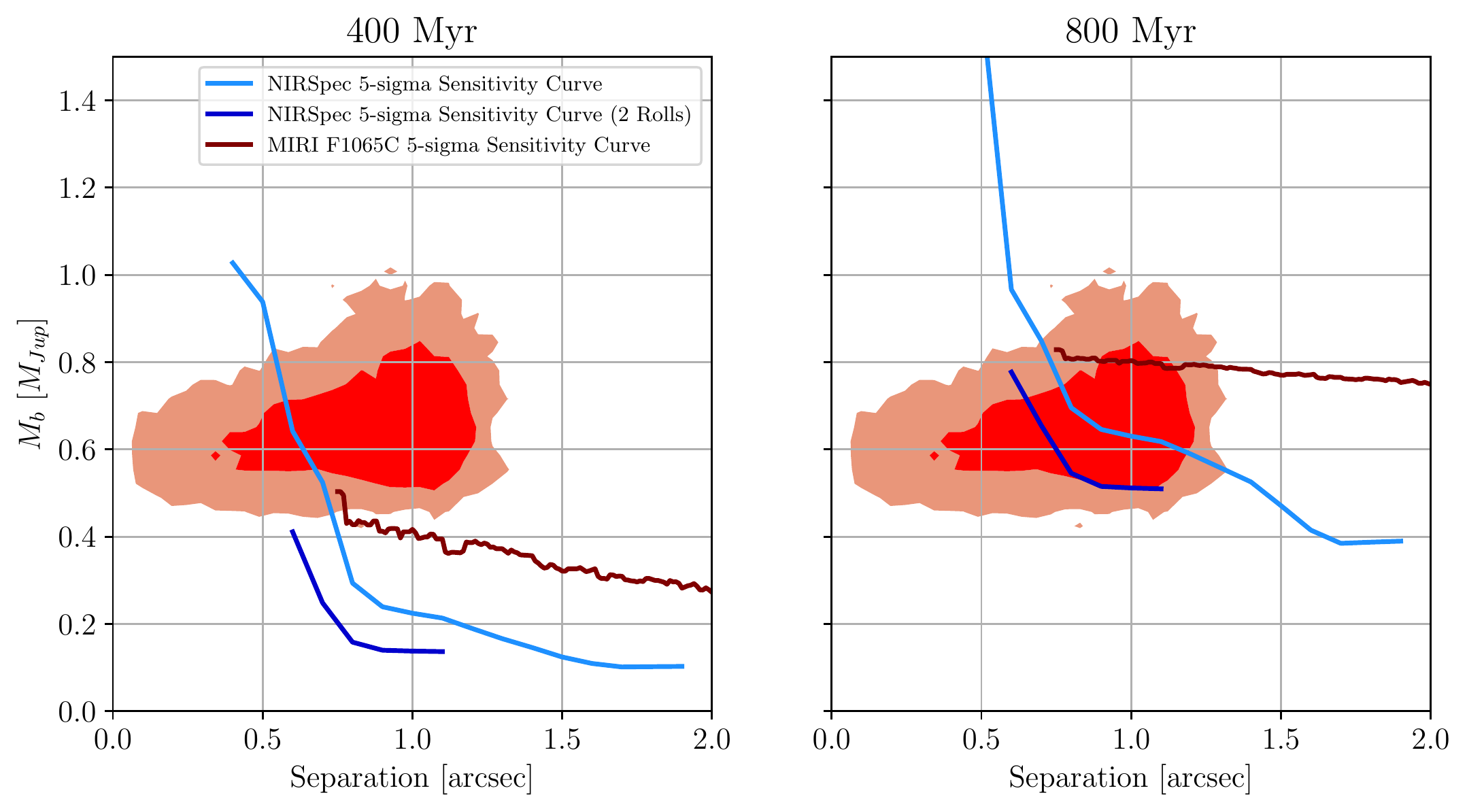}
   \end{tabular}
   \end{center}
   \caption{Expected 5$\sigma$ sensitivity for NIRSpec and MIRI using our data reduction techniques, compared to the expected location and mass of the companion; \textit{red}: 1- and 2-$\sigma$ contours at an epoch close to the expected maximum elongation, i.e. January 2024. The upper sensitivity curve for NIRSpec corresponds to 2 hours of exposure times, while the two-roll case corresponds to a total of 4 hours with a $30^{\circ}$ pupil rotation between two rolls to mitigate the effect of the diffraction spikes of the JWST point spread function. The MIRI simulations require $\sim$75 hours of exposure time to get enough signal-to-noise. These results indicate that NIRSpec is the most sensitive instrument for this science case.
   \label{fig:nirspec_miri}
   } 
\end{figure}

To assess the feasibility of detecting these planets, we simulated NIRSpec observations with the JWST {exposure time calculator} (ETC) and implemented a forward modeling approach similar to \citet{Hoeijmakers2018} and \citet{Ruffio2019} to NIRSpec in which a starlight and a planet model are jointly fitted. The planet model consists of a Sonora atmospheric \citep{Marley2018sonora} modulated by the transmission of the instrument. The same simulation is used to derive the sensitivity of NIRSpec as a function of separation, shown in Fig.~\ref{fig:nirspec_miri}. The JWST ETC does not include many of the likely source of errors that will affect the calibration of the data so the final sensitivity remains uncertain. However, cross correlation techniques are not sensitive to speckle variability and chromaticity, or telescope pointing precision unlike conventional speckle subtraction techniques.
The observations are dominated by the photon noise from the diffracted starlight, so it is critical to minimize the effect of the diffraction spikes in the JWST PSF. They are more than an order of magnitude brighter than the rest of the PSF at a given separation. To avoid chance alignments of the planets with the diffraction spikes, two visits per star with a $30^{\circ}$ pupil rotation can be used to double the average sensitivity of the observation, which is twice as efficient as simply increasing the integration time (Fig.~\ref{fig:nirspec_miri}).
Even in the fastest reading mode and shortest available integration time, the core of the PSF will heavily saturate around $0.6''$ which is limiting the inner working angle. We note that the planets only emit toward the redder part of the band ($4.2\,\mu$m) where the starlight is dimmer, so any wavelength shorter than $4.2\,\mu$m is allowed to saturate with no consequence. Any detector persistence in pixels previously saturated in an earlier dither position will appear like slightly elevated stellar signal, and will naturally be removed by the high pass filtering as if it were speckle noise.

% Figure: NIRSpec Contrast Curves

% **********************************************************************
% **********************************************************************
\subsection{Prospective for a Detection in Reflected Light with the Roman Coronagraph Instrument}
\citet{Carrion-Gonzalez2021} assessed the potential of detecting reflected light from a set of exoplanets in nearby systems with the Roman coronagraph instrument (CGI). In particular, \epsE~stands out as a particularly notable, yet risky target; like most targets for the Roman CGI, it would require thousands of hours to get a detection. They conclude that \epsE~b would be \textit{Roman-accessible} at a probability of 57.99\%, in the optimistic case, and 51.29\% in the pessimistic case. However, they argue that the inclination ``is the key factor affecting the detectability of this planet." Indeed, as it can be seen in their Fig.~C.5, a more edge-on orbit is more favorable to avoiding the outer working angle. Our result for the inclination, $i$ = $78.81_{-22.41}^{\circ+29.34}$, indicates that the orbit should be more favorable for detection, since \citet{Carrion-Gonzalez2021} assume face-on as the most probable orbit.
%
% **********************************************************************
% **********************************************************************
\subsection{Prospects for Ground Based Observatories}
A recent publication by \citet{Pathak2021} presented new observations of \epsE~at 10 $\mu m$ with the VLT/VISIR. They obtained comparable sensitivities to the Keck/NIRC2 results presented here. They claim that a sensitivity to 1 $M_{Jup}$ can be attained with 70 hours of exposure time, assuming an age of 700 Myr and the current setup for the instrument. Unfortunately, the results presented here indicate that the mass is likely lower than 1 $M_{Jup}$. However, there are envisioned ways to upgrading VISIR which would improve its sensitivity at smaller separations \citep{Kasper2019}. 

Although a formidable task for current ground based observatories, imaging \epsE~b should be much easier with future 30-meter class telescopes. {High-contrast imaging at L, M and N bands with METIS at the ELT will essentially be background-limited at the angular separation of \epsE~b \citep{Carlomagno2020}.} METIS is expected to have access to Earth-like planets around $\alpha$~Centauri~A with 5 h of exposure time {N band} \citep{Brandl2021}, which would make \epsE~b detectable in the order of minutes. Similarly, the TMT with its second generation instrument PSI is expected to reach $10^{-8}$ final contrast at 2 $\lambda/D$ \citep{Fitzgerald2019}, well above the sensitivity needed to image \epsE~b.
% **********************************************************************
%
\section{Conclusion}
We combine observations of \epsE~from three different methods: radial velocities spanning three decades, the combined astrometry data from \Hipp~IAD and GRD2, and vortex coronagraph images with Keck/NIRC2. We perform a fit to this data using MCMC and obtain the best constraints to date for the orbital parameters and mass of \epsE~b. Namely, a lower mass posterior with respect to previous analysis \citep{Mawet2019}, $M_b =  0.66_{-0.09}^{+0.12}$~M$_{Jup}$, and new constraints for the inclination, $i$ = $78.81_{-22.41}^{\circ+29.34}$. The new inclination seems to indicate that the planet orbit is not co-planar with the main ring structure, at $i = 34^\circ\pm2^\circ$. Our results are consistent with a small eccentricity, and we improved the accuracy of the time of conjunction to $\sim$81 days. 

These improved constraints translate in a more confident prediction of the position at any epoch as (see Figs.~\ref{fig:contour}~and~\ref{fig:jwstcontour}). We show how this information can be useful when planning the observing strategy, and data reduction, with future missions like the JWST. The JWST is a particularly exciting prospect: we show how NIRSpec could obtain a detection with a reasonable exposure time of just a few hours. More work is expected to be done in this regard. Another exciting landmark for the field of exoplanetary science is the final release of \Gaia's data; we show how, with the expected sensitivity, \Gaia~will most likely have access to a dynamic mass measurement of \epsE~b.

In this paper, we show that the combination of data sets from different observing methods has the power to yield previously inaccessible planetary characteristics from the elusive \epsE~b. As more RV data continues to be collected, and RV facilities and instruments continue to be improved, plus the prospects of Gaia's final data release, and future coronagraph images, the prospects for studying this planet are promising.

\appendix
\section{Radial Velocity Measurements}\label{sec:rv_measurements}
The new RV measurements are shown in Table~\ref{tab:rvs}.

\startlongtable
\begin{deluxetable*}{lrrc}
\tablecaption{ Radial Velocities \label{tab:rvs} }
\tablehead{
  \colhead{Time} & 
  \colhead{RV} & 
  \colhead{RV Unc.} & 
  \colhead{Inst.} \\
  \colhead{(JD)} & 
  \colhead{(m s$^{-1}$)} & 
  \colhead{(m s$^{-1}$)} & 
  \colhead{}
}
\startdata
2457830.7206 & -13.32 & 1.10 & hires$_j$ \\
2457957.0042 & -9.42 & 2.67 & apf \\
2457957.0049 & -18.75 & 2.72 & apf \\
2457957.0055 & -18.07 & 2.86 & apf \\
2457971.9799 & -19.64 & 2.61 & apf \\
2457971.9806 & -21.74 & 2.51 & apf \\
2457971.9814 & -20.58 & 2.53 & apf \\
2457975.9705 & -29.90 & 2.50 & apf \\
2457975.9712 & -26.60 & 2.55 & apf \\
2457975.9720 & -26.59 & 2.45 & apf \\
2457983.0029 & -25.13 & 2.36 & apf \\
2457983.0036 & -28.48 & 2.19 & apf \\
2457983.0044 & -23.69 & 2.33 & apf \\
2457991.9333 & -19.97 & 2.54 & apf \\
2457991.9342 & -19.61 & 2.37 & apf \\
2457991.9350 & -20.31 & 2.42 & apf \\
2457993.0151 & -23.62 & 2.13 & apf \\
2457993.0158 & -30.11 & 2.32 & apf \\
2457993.0166 & -21.83 & 2.05 & apf \\
2458000.1544 & -4.85 & 0.90 & hires$_j$ \\
2458001.1525 & -4.85 & 0.95 & hires$_j$ \\
2458003.1521 & -9.65 & 0.92 & hires$_j$ \\
2458011.8651 & -6.39 & 3.20 & apf \\
2458011.8658 & -14.24 & 3.06 & apf \\
2458011.8665 & -8.81 & 2.98 & apf \\
2458011.9773 & -3.93 & 2.69 & apf \\
2458011.9779 & -9.78 & 2.72 & apf \\
2458011.9786 & -13.28 & 2.60 & apf \\
2458024.9756 & -9.99 & 2.20 & apf \\
2458024.9763 & -10.66 & 2.23 & apf \\
2458024.9770 & -9.56 & 2.46 & apf \\
2458027.8153 & -18.68 & 3.08 & apf \\
2458027.8173 & -16.81 & 3.07 & apf \\
2458027.8191 & -19.40 & 2.79 & apf \\
2458027.9419 & -18.22 & 4.21 & apf \\
2458027.9427 & -18.13 & 2.92 & apf \\
2458027.9434 & -18.38 & 3.45 & apf \\
2458029.9391 & -11.04 & 1.07 & hires$_j$ \\
2458031.9469 & -16.01 & 2.19 & apf \\
2458031.9477 & -15.41 & 2.34 & apf \\
2458031.9485 & -17.23 & 2.22 & apf \\
2458032.9383 & -9.09 & 2.28 & apf \\
2458032.9392 & -6.62 & 2.20 & apf \\
2458032.9400 & -10.73 & 2.25 & apf \\
2458039.8294 & -12.02 & 2.29 & apf \\
2458039.8301 & -12.76 & 2.25 & apf \\
2458039.8309 & -10.73 & 2.37 & apf \\
2458040.9596 & -14.86 & 2.28 & apf \\
2458040.9604 & -15.63 & 2.28 & apf \\
2458040.9612 & -15.16 & 2.30 & apf \\
2458041.9011 & -17.38 & 2.27 & apf \\
2458041.9019 & -20.79 & 2.03 & apf \\
2458041.9027 & -20.04 & 2.10 & apf \\
2458051.7346 & -9.28 & 3.29 & apf \\
2458051.7353 & -13.60 & 3.34 & apf \\
2458051.7360 & -10.66 & 3.23 & apf \\
2458063.8317 & -9.47 & 3.22 & apf \\
2458063.8324 & -18.04 & 3.19 & apf \\
2458063.8332 & -11.10 & 3.49 & apf \\
2458065.9048 & -11.21 & 1.16 & hires$_j$ \\
2458080.8503 & -28.17 & 2.22 & apf \\
2458080.8511 & -28.91 & 2.42 & apf \\
2458080.8519 & -23.24 & 2.28 & apf \\
2458086.7355 & -10.09 & 2.62 & apf \\
2458086.7362 & -17.91 & 2.75 & apf \\
2458086.7369 & -13.81 & 2.65 & apf \\
2458088.6965 & -8.24 & 2.75 & apf \\
2458088.6971 & -8.44 & 2.70 & apf \\
2458088.6978 & -6.73 & 2.93 & apf \\
2458088.8101 & -3.59 & 2.29 & apf \\
2458088.8108 & -7.83 & 2.24 & apf \\
2458088.8116 & -3.13 & 2.32 & apf \\
2458091.9103 & -12.71 & 1.17 & hires$_j$ \\
2458115.6819 & -32.50 & 2.89 & apf \\
2458115.6826 & -37.07 & 2.71 & apf \\
2458115.6832 & -38.80 & 2.76 & apf \\
2458119.7231 & -11.88 & 3.31 & apf \\
2458119.7238 & -9.47 & 2.80 & apf \\
2458119.7245 & -8.95 & 3.07 & apf \\
2458124.8560 & 4.97 & 1.04 & hires$_j$ \\
2458125.6912 & -5.43 & 0.96 & hires$_j$ \\
2458131.6676 & -15.23 & 2.81 & apf \\
2458131.6683 & -3.72 & 2.92 & apf \\
2458131.6690 & -9.99 & 2.99 & apf \\
2458154.7030 & -7.90 & 1.15 & hires$_j$ \\
2458156.6999 & -6.81 & 2.40 & apf \\
2458156.7009 & -6.12 & 2.39 & apf \\
2458156.7018 & -2.55 & 2.40 & apf \\
2458175.6205 & -4.60 & 2.54 & apf \\
2458175.6222 & -9.71 & 2.26 & apf \\
2458175.6236 & -5.39 & 2.21 & apf \\
2458181.7462 & 4.72 & 1.36 & hires$_j$ \\
2458343.0008 & -3.22 & 2.55 & apf \\
2458343.0017 & -4.83 & 2.48 & apf \\
2458343.0025 & -5.67 & 2.54 & apf \\
2458346.1447 & -7.86 & 0.93 & hires$_j$ \\
2458350.1449 & -8.76 & 1.00 & hires$_j$ \\
2458356.9155 & -1.36 & 2.23 & apf \\
2458356.9162 & -4.72 & 2.17 & apf \\
2458356.9170 & 2.84 & 2.08 & apf \\
2458357.9251 & -12.01 & 2.49 & apf \\
2458357.9258 & -16.35 & 2.32 & apf \\
2458357.9266 & -17.44 & 2.35 & apf \\
2458359.9026 & -19.24 & 2.44 & apf \\
2458359.9035 & -25.73 & 2.50 & apf \\
2458359.9043 & -20.22 & 2.51 & apf \\
2458367.0466 & 7.57 & 1.00 & hires$_j$ \\
2458369.0367 & -8.97 & 2.19 & apf \\
2458369.0375 & -7.22 & 2.23 & apf \\
2458369.0382 & -6.04 & 2.14 & apf \\
2458370.0423 & -18.25 & 2.27 & apf \\
2458370.0430 & -18.08 & 2.26 & apf \\
2458370.0437 & -20.25 & 2.41 & apf \\
2458377.0051 & 32.40 & 38.87 & apf \\
2458379.0183 & -4.80 & 2.39 & apf \\
2458379.0189 & -5.50 & 2.50 & apf \\
2458379.0196 & -7.38 & 2.29 & apf \\
2458384.0611 & -16.79 & 1.01 & hires$_j$ \\
2458386.0254 & -9.85 & 1.06 & hires$_j$ \\
2458387.0174 & 0.19 & 1.08 & hires$_j$ \\
2458387.9630 & 3.39 & 2.26 & apf \\
2458387.9637 & 1.27 & 2.42 & apf \\
2458387.9645 & -0.11 & 2.23 & apf \\
2458387.9940 & 8.09 & 1.10 & hires$_j$ \\
2458390.0029 & 1.72 & 2.50 & apf \\
2458390.0042 & 2.69 & 2.37 & apf \\
2458390.0042 & -2.89 & 2.47 & apf \\
2458390.0479 & 6.12 & 0.96 & hires$_j$ \\
2458390.9784 & -3.06 & 2.26 & apf \\
2458390.9792 & 3.35 & 2.41 & apf \\
2458390.9801 & 2.41 & 2.38 & apf \\
2458392.0595 & 2.27 & 1.08 & hires$_j$ \\
2458394.0773 & -7.38 & 1.04 & hires$_j$ \\
2458396.0349 & -10.04 & 1.19 & hires$_j$ \\
2458396.9693 & -3.90 & 1.23 & hires$_j$ \\
2458398.8491 & 4.51 & 3.30 & apf \\
2458398.8499 & -2.61 & 3.38 & apf \\
2458398.8506 & -2.75 & 3.02 & apf \\
2458398.9204 & 0.82 & 2.90 & apf \\
2458398.9211 & 4.59 & 3.02 & apf \\
2458398.9217 & -5.49 & 2.89 & apf \\
2458399.9857 & -0.15 & 2.55 & apf \\
2458399.9866 & 1.95 & 2.51 & apf \\
2458399.9874 & 1.36 & 2.44 & apf \\
2458400.9738 & 0.34 & 2.46 & apf \\
2458400.9746 & -0.60 & 2.37 & apf \\
2458400.9753 & 0.62 & 2.41 & apf \\
2458410.9247 & -12.64 & 2.46 & apf \\
2458410.9254 & -7.35 & 2.31 & apf \\
2458410.9261 & -8.92 & 2.35 & apf \\
2458414.8894 & 9.86 & 2.17 & apf \\
2458414.8902 & 4.71 & 2.10 & apf \\
2458414.8909 & 1.61 & 2.29 & apf \\
2458418.8942 & -21.68 & 2.42 & apf \\
2458418.8950 & -25.37 & 2.40 & apf \\
2458418.8958 & -20.21 & 2.35 & apf \\
2458419.9035 & -13.62 & 2.99 & apf \\
2458419.9045 & -11.57 & 2.79 & apf \\
2458419.9067 & -11.59 & 2.07 & apf \\
2458426.9223 & 10.59 & 1.04 & hires$_j$ \\
2458439.0088 & 7.54 & 1.28 & hires$_j$ \\
2458443.8821 & -10.76 & 1.18 & hires$_j$ \\
2458443.8827 & -6.02 & 1.02 & hires$_j$ \\
2458443.8832 & -10.14 & 1.22 & hires$_j$ \\
2458462.8538 & -2.77 & 1.02 & hires$_j$ \\
2458462.8544 & -1.14 & 1.04 & hires$_j$ \\
2458462.8549 & -2.01 & 1.01 & hires$_j$ \\
2458476.7717 & -3.32 & 1.02 & hires$_j$ \\
2458476.7722 & -2.91 & 1.05 & hires$_j$ \\
2458476.7727 & -3.41 & 0.97 & hires$_j$ \\
2458480.7128 & -4.31 & 2.48 & apf \\
2458480.7139 & 2.18 & 2.60 & apf \\
2458480.7150 & -0.89 & 2.54 & apf \\
2458487.6420 & -4.39 & 2.37 & apf \\
2458487.6427 & -4.97 & 2.28 & apf \\
2458487.6434 & -6.74 & 2.41 & apf \\
2458488.6304 & -16.61 & 2.84 & apf \\
2458488.6311 & -17.42 & 2.53 & apf \\
2458488.6317 & -15.66 & 2.40 & apf \\
2458490.7571 & 2.75 & 1.04 & hires$_j$ \\
2458490.7577 & 0.01 & 1.05 & hires$_j$ \\
2458490.7582 & -1.50 & 1.06 & hires$_j$ \\
2458532.7200 & -4.04 & 1.12 & hires$_j$ \\
2458568.7106 & -7.45 & 1.08 & hires$_j$ \\
2458568.7112 & -8.92 & 1.17 & hires$_j$ \\
2458568.7117 & -7.60 & 1.12 & hires$_j$ \\
2458569.7117 & -7.98 & 1.00 & hires$_j$ \\
2458569.7123 & -7.88 & 1.10 & hires$_j$ \\
2458569.7128 & -8.14 & 1.02 & hires$_j$ \\
2458714.1492 & -9.22 & 1.08 & hires$_j$ \\
2458715.1508 & -1.96 & 0.93 & hires$_j$ \\
2458716.1499 & -3.73 & 0.89 & hires$_j$ \\
2458723.1549 & 8.30 & 0.96 & hires$_j$ \\
2458724.1550 & 1.85 & 0.93 & hires$_j$ \\
2458732.0204 & -7.99 & 2.41 & apf \\
2458732.0230 & -9.53 & 2.27 & apf \\
2458732.0252 & -7.23 & 2.32 & apf \\
2458733.1529 & 0.70 & 1.04 & hires$_j$ \\
2458744.1572 & -2.81 & 1.16 & hires$_j$ \\
2458746.8760 & -2.41 & 2.33 & apf \\
2458746.8768 & 3.22 & 2.40 & apf \\
2458746.8777 & 1.49 & 2.39 & apf \\
2458746.8785 & 2.54 & 2.26 & apf \\
2458747.8382 & 6.14 & 2.72 & apf \\
2458747.8391 & 0.56 & 2.66 & apf \\
2458747.8401 & 2.39 & 2.84 & apf \\
2458747.8410 & 5.68 & 2.69 & apf \\
2458749.8802 & 1.21 & 2.64 & apf \\
2458749.8812 & 5.31 & 2.52 & apf \\
2458749.8822 & 3.23 & 2.64 & apf \\
2458752.9836 & -15.60 & 2.07 & apf \\
2458752.9844 & -11.82 & 2.24 & apf \\
2458752.9852 & -10.04 & 2.22 & apf \\
2458752.9860 & -12.86 & 2.27 & apf \\
2458765.8514 & 0.59 & 2.51 & apf \\
2458765.8523 & -1.54 & 2.25 & apf \\
2458765.8533 & -5.55 & 2.65 & apf \\
2458776.9385 & -2.09 & 1.14 & hires$_j$ \\
2458794.9139 & 10.33 & 1.06 & hires$_j$ \\
2458795.9719 & 11.51 & 0.96 & hires$_j$ \\
2458796.9731 & 8.51 & 1.17 & hires$_j$ \\
2458797.9759 & 3.21 & 1.01 & hires$_j$ \\
2458798.8832 & 15.73 & 3.20 & apf \\
2458798.8838 & 15.91 & 4.46 & apf \\
2458798.8844 & 10.47 & 5.65 & apf \\
2458798.9255 & 9.65 & 1.05 & hires$_j$ \\
2458800.7426 & 0.40 & 8.44 & apf \\
2458800.7431 & -11.34 & 10.11 & apf \\
2458800.7437 & 0.37 & 8.28 & apf \\
2458802.9074 & -1.55 & 1.04 & hires$_j$ \\
2458819.9272 & -1.62 & 1.23 & hires$_j$ \\
2458819.9278 & 2.20 & 1.17 & hires$_j$ \\
2458819.9285 & -1.86 & 1.25 & hires$_j$ \\
2458880.7768 & 14.95 & 1.09 & hires$_j$ \\
2458905.7045 & 7.52 & 0.93 & hires$_j$ \\
2458906.7048 & 9.72 & 1.01 & hires$_j$ \\
2458907.7049 & 11.37 & 1.05 & hires$_j$ \\
2459064.1408 & 0.66 & 0.88 & hires$_j$ \\
2459067.1418 & 4.26 & 0.92 & hires$_j$ \\
2459069.0914 & 16.14 & 0.91 & hires$_j$ \\
2459079.1461 & 6.80 & 0.97 & hires$_j$ \\
2459088.1435 & -0.35 & 0.83 & hires$_j$ \\
2459089.1488 & -0.30 & 1.06 & hires$_j$ \\
2459090.1506 & 5.99 & 0.85 & hires$_j$ \\
2459091.1516 & 12.06 & 0.89 & hires$_j$ \\
2459117.1542 & 16.88 & 1.01 & hires$_j$ \\
2459118.1551 & 17.29 & 0.96 & hires$_j$ \\
2459120.0833 & 16.54 & 0.94 & hires$_j$ \\ 
\enddata
%\tablecomments{  }
\end{deluxetable*}

\section{Hipparcos IAD Measurements}\label{sec:IADtable}
The full list of IAD measurements of \epsE~are shown in Table~\ref{tab:data_IAD}.

% \begin{longtable*}{ccc}
\startlongtable
\begin{deluxetable*}{ccc}
\tablecaption{ \Hipp~IAD Measurements \label{tab:data_IAD} }
\tablehead{
  \colhead{Time} & 
  \colhead{RA} & 
  \colhead{Dec.}\\
  \colhead{(yr)} & 
  \colhead{($^\circ$)} & 
  \colhead{($^\circ$)}
}
\startdata
1990.036 & 52.5116405005777 & -9.4583497858903 \\
1990.036 & 52.5116401584946 & -9.4583496322689 \\
1990.036 & 52.5116403587071 & -9.4583497221005 \\
1990.036 & 52.5116403865140 & -9.4583497347661 \\
1990.165 & 52.5115962411255 & -9.4583180708649 \\
1990.165 & 52.5115959467574 & -9.4583177566427 \\
1990.165 & 52.5115960096074 & -9.4583178233896 \\
1990.537 & 52.5116469916317 & -9.4582724641828 \\
1990.537 & 52.5116469841479 & -9.4582725080033 \\
1990.598 & 52.5116419591556 & -9.4582840685775 \\
1990.598 & 52.5116414487723 & -9.4582837747986 \\
1990.598 & 52.5116417640389 & -9.4582839565381 \\
1990.598 & 52.5116420312751 & -9.4582841103275 \\
1990.986 & 52.5114013929774 & -9.4583510281954 \\
1990.986 & 52.5114013393019 & -9.4583509989263 \\
1990.986 & 52.5114015402335 & -9.4583511001388 \\
1990.986 & 52.5114012050044 & -9.4583509320788 \\
1990.986 & 52.5114009365369 & -9.4583507920843 \\
1990.986 & 52.5114013955396 & -9.4583510292821 \\
1990.986 & 52.5114010690108 & -9.4583508624671 \\
1991.033 & 52.5113714176233 & -9.4583447128771 \\
1991.033 & 52.5113714135922 & -9.4583447403729 \\
1991.033 & 52.5113713999200 & -9.4583448169388 \\
1991.033 & 52.5113714029650 & -9.4583448033679 \\
1991.211 & 52.5113225169863 & -9.4582994758077 \\
1991.211 & 52.5113226736580 & -9.4582992298002 \\
1991.211 & 52.5113224771832 & -9.4582995327838 \\
1991.211 & 52.5113225766121 & -9.4582993787588 \\
1991.211 & 52.5113225353019 & -9.4582994446619 \\
1991.211 & 52.5113226173269 & -9.4582993157660 \\
1991.211 & 52.5113225487930 & -9.4582994236099 \\
1991.211 & 52.5113226206707 & -9.4582993113122 \\
1991.233 & 52.5113235973147 & -9.4582931095657 \\
1991.233 & 52.5113235887250 & -9.4582928458202 \\
1991.233 & 52.5113235967572 & -9.4582930845733 \\
1991.233 & 52.5113236047697 & -9.4582933483541 \\
1991.233 & 52.5113235890592 & -9.4582926401783 \\
1991.233 & 52.5113235894297 & -9.4582926262907 \\
1991.233 & 52.5113235941072 & -9.4582930012824 \\
1991.233 & 52.5113236004370 & -9.4582932122974 \\
1991.487 & 52.5113727458625 & -9.4582615280949 \\
1991.487 & 52.5113727288772 & -9.4582614578977 \\
1991.487 & 52.5113727022689 & -9.4582613500171 \\
1991.487 & 52.5113727581897 & -9.4582615765491 \\
1991.487 & 52.5113728058925 & -9.4582617679102 \\
1991.487 & 52.5113727896355 & -9.4582617118418 \\
1991.524 & 52.5113756856158 & -9.4582652690679 \\
1991.524 & 52.5113756602019 & -9.4582652578526 \\
1991.617 & 52.5113670292940 & -9.4582828689332 \\
1991.617 & 52.5113670158256 & -9.4582828932237 \\
1991.618 & 52.5113670300430 & -9.4582826870404 \\
1991.618 & 52.5113667689413 & -9.4582831584554 \\
1991.714 & 52.5113300285460 & -9.4583083881100 \\
1991.714 & 52.5113300175102 & -9.4583083656777 \\
1991.714 & 52.5113301324919 & -9.4583086000961 \\
1991.714 & 52.5113300055516 & -9.4583083406066 \\
1991.714 & 52.5113299595752 & -9.4583082455852 \\
1991.714 & 52.5113300402794 & -9.4583084132933 \\
1991.715 & 52.5113293922848 & -9.4583084348328 \\
1991.715 & 52.5113293727876 & -9.4583083887744 \\
1991.715 & 52.5113294670004 & -9.4583085961528 \\
1992.062 & 52.5110857608807 & -9.4583343386895 \\
1992.062 & 52.5110853444093 & -9.4583341343475 \\
1992.062 & 52.5110851647440 & -9.4583340464981 \\
1992.062 & 52.5110856265134 & -9.4583342719861 \\
1992.062 & 52.5110853695640 & -9.4583341461406 \\
1992.139 & 52.5110586465189 & -9.4583148046891 \\
1992.139 & 52.5110585719244 & -9.4583149218441 \\
1992.139 & 52.5110585808650 & -9.4583149077813 \\
1992.139 & 52.5110585956311 & -9.4583148842530 \\
1992.189 & 52.5110520744159 & -9.4583002913271 \\
1992.189 & 52.5110521033931 & -9.4583003354390 \\
1992.189 & 52.5110521308103 & -9.4583003772504 \\
1992.189 & 52.5110519452936 & -9.4583000936515 \\
1992.563 & 52.5111043447049 & -9.4582660986126 \\
1992.563 & 52.5111044409491 & -9.4582658056901 \\
1992.563 & 52.5111043696635 & -9.4582660220212 \\
1992.563 & 52.5111043689974 & -9.4582660247248 \\
\enddata
% \tablecomments{  }
\end{deluxetable*}
% \end{longtable*}

\section{Using the Kernel Density Estimator }\label{sec:kde}
\begin{figure}[t!]
   \begin{center}
   \begin{tabular}{c} %% 
   \includegraphics[height=10.0cm]{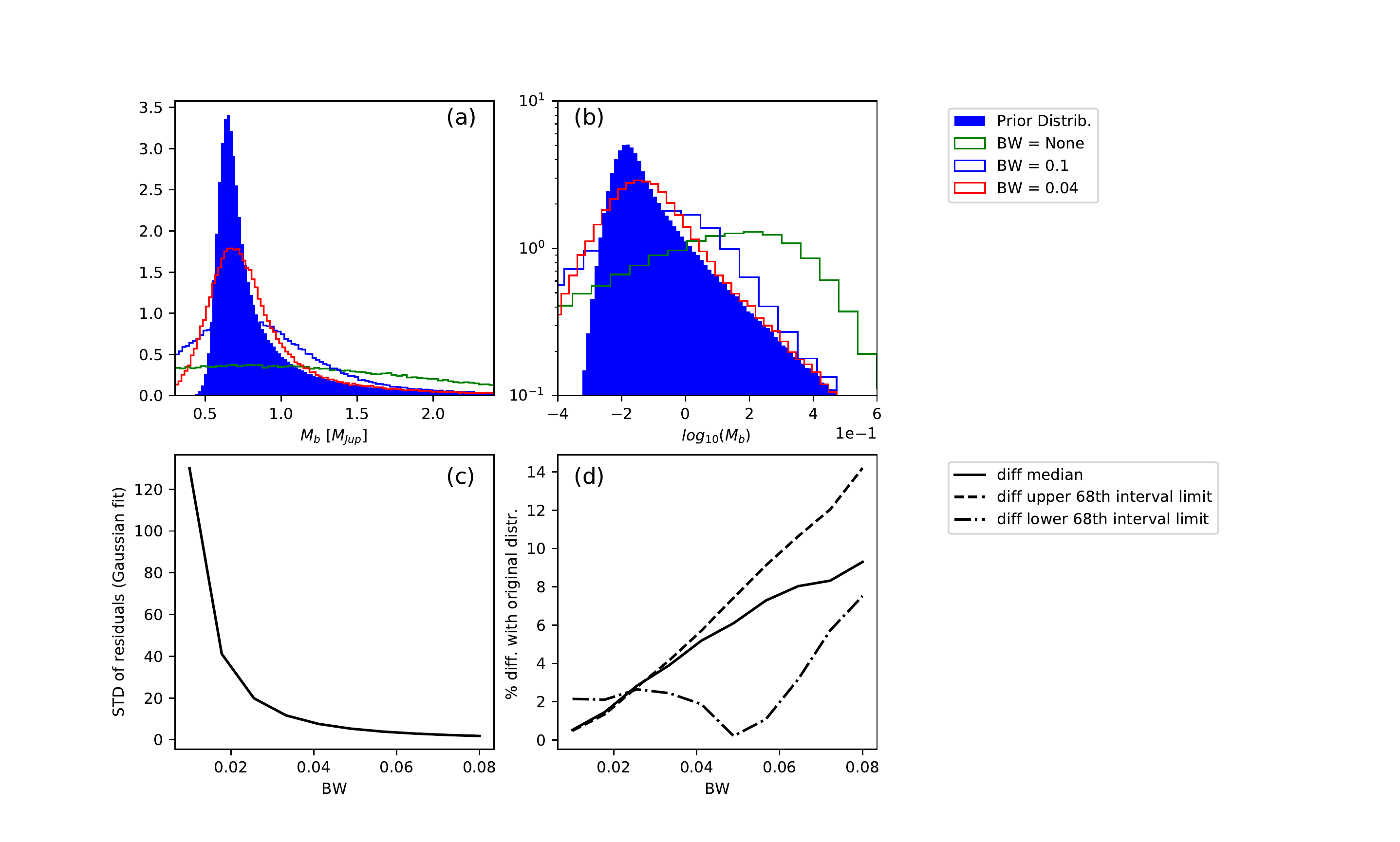}
   \end{tabular}
   \end{center}
   \caption{
   \label{fig:kde_bw}
   (a) Mass-histogram comparing three different KDE bandwidths to fit the 6-correlated parameters from the RV fit. The blue bar histogram represents the distribution from the RV fit, the step histograms are KDE fit to that distribution for different bandwidths. (b) The residuals in prior probability to a Gaussian fit versus the KDE bandwidth. For a set of reasonable SMAs, we compute the prior probabilities, and we fit a Gaussian; the narrower the bandwidth, the more the spurious effects of the RV posterior sampling affect the KDE fit. (c) Difference between the KDE fit and the original prior distribution for the mass distribution. The narrower the bandwidth, the closer to the original distribution. (d) Difference of the confidence intervals with respect to the original distribution. The narrower the bandwidth the more the upper and lower bounds are similar to the original distribution. A balance, thus, needs to be found between a small enough bandwidth so that the upper and lower bounds are well reproduced (see (d)), but not as small as to begin an undersmoothing effect (see (c)). 
   } 
\end{figure}

The bandwidth is in practice associated with the smoothing of the kernel, and has to be carefully tuned to (1) avoid data artifacts caused by undersmoothing, and (2) retain the distribution tail information and skewed bounds limits that could be lost by oversmoothing. In Fig.~\ref{fig:kde_bw}(a) we illustrate these effects and the importance of bandwidth selection. The value of the optimal bandwidth is, however, heavily dependent on the data; a higher amount of data points allows for a more aggressive bandwidth, i.e. lower bandwidth, and a low number of data points is more susceptible to a spurious ridged distribution. 

A way of choosing the KDE bandwidth is:
1. Pick an acceptable change in the median and 68th interval limits of the KDE fit w.r.t the actual posterior distribution. This will set a maximum acceptable bandwidth.
2. Pick an acceptable variation of the log-prior probability when evaluating the priors for a SMA around the prior median SMA. This will set a minimum acceptable bandwidth.
For this we will loop over a set of bandwidths computing the median and 68th interval limits, and for each bandwidth we will compute the variation of log-prior with a set of SMAs around the prior median SMA: we will fit a Gaussian and the standard deviation of the residuals to the fit will be our cost function.

% Figure: KDE stuff 

%%%%%%%%%%%%%%%%%%%%%%%%%%%%%%%%%%%%%%%%%%%%%%%%%%%%%
\section{Perturbed Orbit Solution}\label{sec:app_perturbed}
\begin{figure}[t!]
   \begin{center}
   \begin{tabular}{c} %% 
   \includegraphics[height=11.0cm]{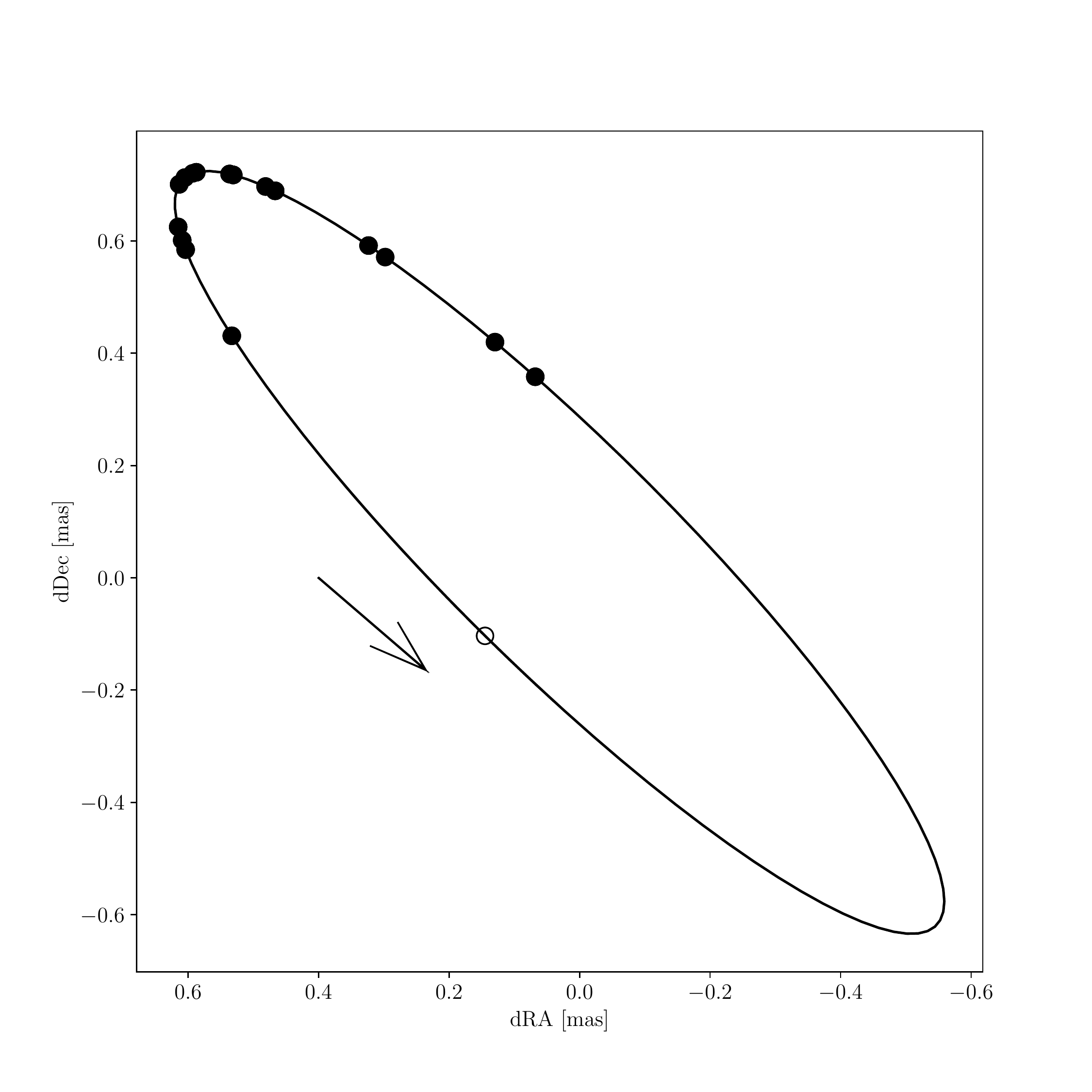}
   \end{tabular}
   \end{center}
   \caption{
   \label{fig:perturbed}
   \epsE's~perturbed orbit caused by the presence of the companion. Filled circles indicate the estimated position of the star at the \Hipp~IAD epochs, the empty circle indicates the \Gaia~DR2 epoch. The arrow indicates the direction of motion.} 
\end{figure}

The perturbed orbit for the case of 800 Myr is shown in Fig.~\ref{fig:perturbed}; the perturbation size is $\alpha_A$ = 0.89 mas. Overplotted are the estimated positions of \epsE~ for the \Hipp~and \Gaia~epochs; \Hipp~covers $\sim$35\% of the orbit, \Gaia~epoch is taken as 2015.5.

%%%%%%%%%%%%%%%%%%%%%%%%%%%%%%%%%%%%%%%%%%%%%%%%%%%%%
\section{Corner Plot}\label{sec:app_cornerplot}
\begin{figure}[t!]
   \begin{center}
   \begin{tabular}{c} %% 
   \includegraphics[height=16.0cm]{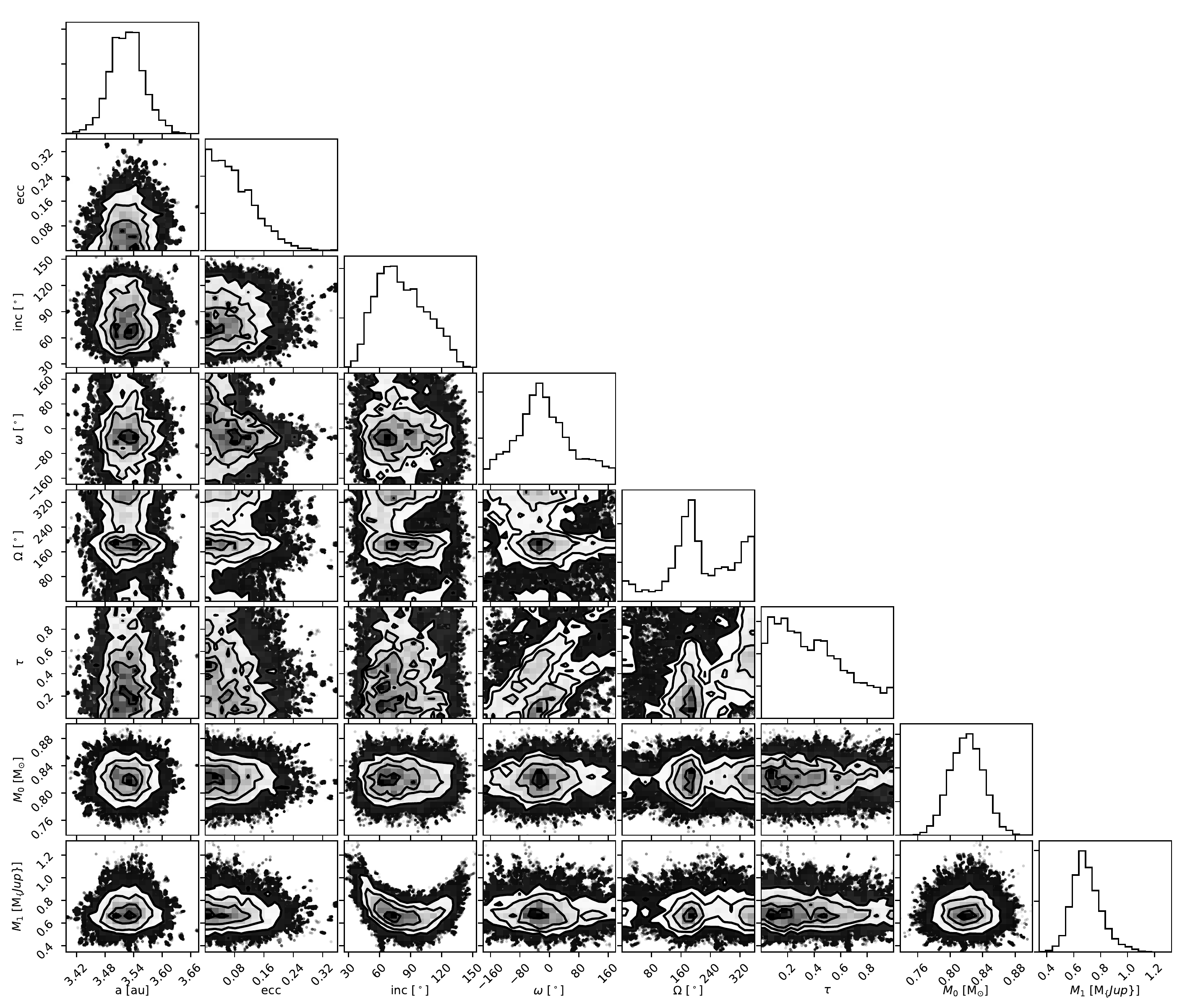}
   \end{tabular}
   \end{center}
   \caption{
   \label{fig:corner_plot_full}
   Posteriors of the orbital parameters and masses.} 
\end{figure}

In Fig.~\ref{fig:corner_plot_full} the full corner plot is shown, with the posterior distributions and their correlation for the six orbital parameters and the masses. These posteriors correspond to the MCMC run for the fit to the RV, astrometry, and direct imaging data assuming an age of 800 Myr.

\acknowledgments % equivalent to 
This work was partially supported by the National Science Foundation AST-ATI Grant 1710210. The material is based upon work supported by NASA under award No. 80NSSC20K0624.
J.J.W. is supported by the Heising-Simons Foundation 51 Pegasi b Fellowship. J.J.W. thanks Rob de Rosa and Eric Nielsen for helpful discussions.
P. D. is supported by a National Science Foundation (NSF) Astronomy and Astrophysics Postdoctoral Fellowship under award AST-1903811. M.R. is supported by the National Science Foundation Graduate Research Fellowship Program under Grant Number DGE-1752134
Part of this work has received funding from
the European Research Council (ERC) under the European Union's Horizon
2020 research and innovation programme (grant agreement No. 819155), and
by the Wallonia-Brussels Federation (grant for Concerted Research Actions). J.L-S. thanks Christopher Mendillo for helpful discussions.

\software{pyKLIP \citep{Wang2015}, Scipy\citep{Jones2001}, \orbitize \citep{Blunt2020}, emcee \citep{Foreman-Mackey2013}, corner.py \citep{Foreman-Mackey2016}, \texttt{RVSearch} (Rosenthal et al. 2021), Astropy \citep{Astropy},
Matplotlib \citep{Matplotlib}}

\bibliography{report}{}

\begin{thebibliography}{}
\expandafter\ifx\csname natexlab\endcsname\relax\def\natexlab#1{#1}\fi
\providecommand{\url}[1]{\href{#1}{#1}}
\providecommand{\dodoi}[1]{doi:~\href{http://doi.org/#1}{\nolinkurl{#1}}}
\providecommand{\doeprint}[1]{\href{http://ascl.net/#1}{\nolinkurl{http://ascl.net/#1}}}
\providecommand{\doarXiv}[1]{\href{https://arxiv.org/abs/#1}{\nolinkurl{https://arxiv.org/abs/#1}}}

\bibitem[{{Astropy Collaboration} {et~al.}(2013){Astropy Collaboration},
  {Robitaille}, {Tollerud}, {Greenfield}, {Droettboom}, {Bray}, {Aldcroft},
  {Davis}, {Ginsburg}, {Price-Whelan}, {Kerzendorf}, {Conley}, {Crighton},
  {Barbary}, {Muna}, {Ferguson}, {Grollier}, {Parikh}, {Nair}, {Unther},
  {Deil}, {Woillez}, {Conseil}, {Kramer}, {Turner}, {Singer}, {Fox}, {Weaver},
  {Zabalza}, {Edwards}, {Azalee Bostroem}, {Burke}, {Casey}, {Crawford},
  {Dencheva}, {Ely}, {Jenness}, {Labrie}, {Lim}, {Pierfederici}, {Pontzen},
  {Ptak}, {Refsdal}, {Servillat}, \& {Streicher}}]{Astropy}
{Astropy Collaboration}, {Robitaille}, T.~P., {Tollerud}, E.~J., {et~al.} 2013,
  \aap, 558, A33, \dodoi{10.1051/0004-6361/201322068}

\bibitem[{{Aumann}(1985)}]{Aumann1985}
{Aumann}, H.~H. 1985, \pasp, 97, 885, \dodoi{10.1086/131620}

\bibitem[{Backman {et~al.}(2008)Backman, Marengo, Stapelfeldt, Su, Wilner,
  Dowell, Watson, Stansberry, Rieke, Megeath, Fazio, \& Werner}]{Backman2008}
Backman, D., Marengo, M., Stapelfeldt, K., {et~al.} 2008, The Astrophysical
  Journal, 690, 1522, \dodoi{10.1088/0004-637x/690/2/1522}

\bibitem[{Baines \& Armstrong(2011)}]{Baines2012}
Baines, E.~K., \& Armstrong, J.~T. 2011, The Astrophysical Journal, 744, 138,
  \dodoi{10.1088/0004-637x/744/2/138}

\bibitem[{{Baraffe} {et~al.}(2003){Baraffe}, {Chabrier}, {Barman}, {Allard}, \&
  {Hauschildt}}]{Baraffe2003}
{Baraffe}, I., {Chabrier}, G., {Barman}, T.~S., {Allard}, F., \& {Hauschildt},
  P.~H. 2003, \aap, 402, 701, \dodoi{10.1051/0004-6361:20030252}

\bibitem[{{Barman} {et~al.}(2015){Barman}, {Konopacky}, {Macintosh}, \&
  {Marois}}]{Barman2015}
{Barman}, T.~S., {Konopacky}, Q.~M., {Macintosh}, B., \& {Marois}, C. 2015,
  \apj, 804, 61, \dodoi{10.1088/0004-637X/804/1/61}

\bibitem[{{Beichman} {et~al.}(2020){Beichman}, {Ygouf}, {Llop Sayson}, {Mawet},
  {Yung}, {Choquet}, {Kervella}, {Boccaletti}, {Belikov}, {Lissauer},
  {Quarles}, {Lagage}, {Dicken}, {Hu}, {Mennesson}, {Ressler}, {Serabyn},
  {Krist}, {Bendek}, {Leisenring}, \& {Pueyo}}]{Beichman2020}
{Beichman}, C., {Ygouf}, M., {Llop Sayson}, J., {et~al.} 2020, \pasp, 132,
  015002, \dodoi{10.1088/1538-3873/ab5066}

\bibitem[{Benedict {et~al.}(2006)Benedict, McArthur, Gatewood, Nelan, Cochran,
  Hatzes, Endl, Wittenmyer, Baliunas, Walker, Yang, K{\"u}rster, Els, \&
  Paulson}]{Benedict2006}
Benedict, G.~F., McArthur, B.~E., Gatewood, G., {et~al.} 2006, The Astronomical
  Journal, 132, 2206, \dodoi{10.1086/508323}

\bibitem[{{Blunt} {et~al.}(2019){Blunt}, {Endl}, {Weiss}, {Cochran}, {Howard},
  {MacQueen}, {Fulton}, {Henry}, {Johnson}, {Kosiarek}, {Lawson}, {Macintosh},
  {Mills}, {Nielsen}, {Petigura}, {Schneider}, {Vanderburg}, {Wisniewski},
  {Wittenmyer}, {Brugamyer}, {Caldwell}, {Cochran}, {Hatzes}, {Hirsch},
  {Isaacson}, {Robertson}, {Roy}, \& {Shen}}]{Blunt2019}
{Blunt}, S., {Endl}, M., {Weiss}, L.~M., {et~al.} 2019, \aj, 158, 181,
  \dodoi{10.3847/1538-3881/ab3e63}

\bibitem[{{Blunt} {et~al.}(2020){Blunt}, {Wang}, {Angelo}, {Ngo}, {Cody}, {De
  Rosa}, {Graham}, {Hirsch}, {Nagpal}, {Nielsen}, {Pearce}, {Rice}, \&
  {Tejada}}]{Blunt2020}
{Blunt}, S., {Wang}, J.~J., {Angelo}, I., {et~al.} 2020, \aj, 159, 89,
  \dodoi{10.3847/1538-3881/ab6663}

\bibitem[{Boccaletti {et~al.}(2015)Boccaletti, Lagage, Baudoz, Beichman,
  Bouchet, Cavarroc, Dubreuil, Glasse, Glauser, Hines, \&
  et~al.}]{Boccaletti2015}
Boccaletti, A., Lagage, P.-O., Baudoz, P., {et~al.} 2015, Publications of the
  Astronomical Society of the Pacific, 127, 633–645, \dodoi{10.1086/682256}

\bibitem[{{Bond} {et~al.}(2020){Bond}, {Cetre}, {Lilley}, {Wizinowich},
  {Mawet}, {Chun}, {Wetherell}, {Jacobson}, {Lockhart}, {Warmbier}, {Ragland},
  {Alvarez}, {Guyon}, {Goebel}, {Delorme}, {Jovanovic}, {Hall}, {Wallace},
  {Taheri}, {Plantet}, \& {Chambouleyron}}]{Bond2020}
{Bond}, C.~Z., {Cetre}, S., {Lilley}, S., {et~al.} 2020, Journal of
  Astronomical Telescopes, Instruments, and Systems, 6, 039003,
  \dodoi{10.1117/1.JATIS.6.3.039003}

\bibitem[{{Booth} {et~al.}(2017){Booth}, {Dent}, {Jord{\'a}n}, {Lestrade},
  {Hales}, {Wyatt}, {Casassus}, {Ertel}, {Greaves}, {Kennedy}, {Matr{\`a}},
  {Augereau}, \& {Villard}}]{Booth2017}
{Booth}, M., {Dent}, W. R.~F., {Jord{\'a}n}, A., {et~al.} 2017, \mnras, 469,
  3200, \dodoi{10.1093/mnras/stx1072}

\bibitem[{{Brandl} {et~al.}(2021){Brandl}, {Bettonvil}, {van Boekel},
  {Glauser}, {Quanz}, {Absil}, {Amorim}, {Feldt}, {Glasse}, {G{\"u}del}, {Ho},
  {Labadie}, {Meyer}, {Pantin}, {van Winckel}, \& {METIS
  Consortium}}]{Brandl2021}
{Brandl}, B., {Bettonvil}, F., {van Boekel}, R., {et~al.} 2021, The Messenger,
  182, 22, \dodoi{10.18727/0722-6691/5218}

\bibitem[{{Brandt}(2021)}]{Brandt2021}
{Brandt}, T.~D. 2021, arXiv e-prints, arXiv:2105.11662.
\newblock \doarXiv{2105.11662}

\bibitem[{{Carlomagno} {et~al.}(2020){Carlomagno}, {Delacroix}, {Absil},
  {Cantalloube}, {Orban de Xivry}, {Pathak}, {Agocs}, {Bertram}, {Brandl},
  {Burtscher}, {Feldt}, {Glauser}, {Hippler}, {Kenworthy}, {Stuik}, \& {van
  Boekel}}]{Carlomagno2020}
{Carlomagno}, B., {Delacroix}, C., {Absil}, O., {et~al.} 2020, Journal of
  Astronomical Telescopes, Instruments, and Systems, 6, 035005,
  \dodoi{10.1117/1.JATIS.6.3.035005}

\bibitem[{{Carri{\'o}n-Gonz{\'a}lez} {et~al.}(2021){Carri{\'o}n-Gonz{\'a}lez},
  {Garc{\'\i}a Mu{\~n}oz}, {Cabrera}, {Csizmadia}, {Santos}, \&
  {Rauer}}]{Carrion-Gonzalez2021}
{Carri{\'o}n-Gonz{\'a}lez}, {\'O}., {Garc{\'\i}a Mu{\~n}oz}, A., {Cabrera}, J.,
  {et~al.} 2021, arXiv e-prints, arXiv:2104.04296.
\newblock \doarXiv{2104.04296}

\bibitem[{{Cox}(2000)}]{Cox2000}
{Cox}, A.~N. 2000, {Allen's astrophysical quantities}

\bibitem[{Dawson {et~al.}(2011)Dawson, Murray-Clay, \& Fabrycky}]{Dawson2011}
Dawson, R.~I., Murray-Clay, R.~A., \& Fabrycky, D.~C. 2011, The Astrophysical
  Journal, 743, L17, \dodoi{10.1088/2041-8205/743/1/l17}

\bibitem[{{De Rosa} {et~al.}(2019{\natexlab{a}}){De Rosa}, Esposito, Hirsch,
  Nielsen, Marley, Kalas, Wang, \& Macintosh}]{DeRosa2019}
{De Rosa}, R.~J., Esposito, T.~M., Hirsch, L.~A., {et~al.} 2019{\natexlab{a}},
  The Astronomical Journal, 158, 225, \dodoi{10.3847/1538-3881/ab4c9b}

\bibitem[{{De Rosa} {et~al.}(2019{\natexlab{b}}){De Rosa}, Nielsen, Wang,
  Ammons, Duch{\^{e}}ne, Macintosh, Rameau, Bailey, Barman, Bulger, Chilcote,
  Cotten, Doyon, Esposito, Fitzgerald, Follette, Gerard, Goodsell, Graham,
  Greenbaum, Hibon, Hom, Hung, Ingraham, Kalas, Konopacky, Larkin, Maire,
  Marchis, Marley, Marois, Metchev, Millar-Blanchaer, Oppenheimer, Palmer,
  Patience, Perrin, Poyneer, Pueyo, Rajan, Rantakyr{\"o}, Ren, Ruffio,
  Savransky, Schneider, Sivaramakrishnan, Song, Soummer, Tallis, Thomas,
  Wallace, Ward-Duong, Wiktorowicz, \& Wolff}]{DeRosa2020}
{De Rosa}, R.~J., Nielsen, E.~L., Wang, J.~J., {et~al.} 2019{\natexlab{b}}, The
  Astronomical Journal, 159, 1

\bibitem[{{Ducati}(2002)}]{Ducati2002}
{Ducati}, J.~R. 2002, VizieR Online Data Catalog, 2237

\bibitem[{{Ertel} {et~al.}(2018){Ertel}, {Defr{\`e}re}, {Hinz}, {Mennesson},
  {Kennedy}, {Danchi}, {Gelino}, {Hill}, {Hoffmann}, {Rieke}, {Shannon},
  {Spalding}, {Stone}, {Vaz}, {Weinberger}, {Willems}, {Absil}, {Arbo},
  {Bailey}, {Beichman}, {Bryden}, {Downey}, {Durney}, {Esposito}, {Gaspar},
  {Grenz}, {Haniff}, {Leisenring}, {Marion}, {McMahon}, {Millan-Gabet},
  {Montoya}, {Morzinski}, {Pinna}, {Power}, {Puglisi}, {Roberge}, {Serabyn},
  {Skemer}, {Stapelfeldt}, {Su}, {Vaitheeswaran}, \& {Wyatt}}]{Ertel2018}
{Ertel}, S., {Defr{\`e}re}, D., {Hinz}, P., {et~al.} 2018, \aj, 155, 194,
  \dodoi{10.3847/1538-3881/aab717}

\bibitem[{{Fitzgerald} {et~al.}(2019){Fitzgerald}, {Bailey}, {Baranec},
  {Batalha}, {Benneke}, {Beichman}, {Brandt}, {Chilcote}, {Chun}, {Crossfield},
  {Currie}, {Davis}, {Dekany}, {Delorme}, {Dong}, {Doyon}, {Dressing},
  {Echeverri}, {Fortney}, {Frazin}, {Guyon}, {Hashimoto}, {Hillenbrand},
  {Hinz}, {Howard}, {Jensen-Clem}, {Jovanovic}, {Kawahara}, {Knutson},
  {Konopacky}, {Kotani}, {Lafreni{\`e}re}, {Liu}, {Lozi}, {Lu}, {Males},
  {Marley}, {Marois}, {Mawet}, {Mazin}, {Millar-Blanchaer}, {Mondal},
  {Murakami}, {Murray-Clay}, {Narita}, {Pezzato}, {Pyo}, {Roberts}, {Ruane},
  {Sallum}, {Serabyn}, {Shields}, {Simard}, {Skemer}, {Stelter}, {Tamura},
  {Troy}, {Vasisht}, {Wallace}, {Wang}, {Wang}, \& {Wright}}]{Fitzgerald2019}
{Fitzgerald}, M., {Bailey}, V., {Baranec}, C., {et~al.} 2019, in Bulletin of
  the American Astronomical Society, Vol.~51, 251

\bibitem[{Foreman-Mackey(2016)}]{Foreman-Mackey2016}
Foreman-Mackey, D. 2016, Journal of Open Source Software, 1, 24,
  \dodoi{10.21105/joss.00024}

\bibitem[{Foreman-Mackey {et~al.}(2013)Foreman-Mackey, Hogg, Lang, \&
  Goodman}]{Foreman-Mackey2013}
Foreman-Mackey, D., Hogg, D.~W., Lang, D., \& Goodman, J. 2013, Publications of
  the Astronomical Society of the Pacific, 125, 306–312,
  \dodoi{10.1086/670067}

\bibitem[{{Fulton} {et~al.}(2018){Fulton}, {Petigura}, {Blunt}, \&
  {Sinukoff}}]{Fulton2018}
{Fulton}, B.~J., {Petigura}, E.~A., {Blunt}, S., \& {Sinukoff}, E. 2018, \pasp,
  130, 044504, \dodoi{10.1088/1538-3873/aaaaa8}

\bibitem[{{Gaia Collaboration} {et~al.}(2018){Gaia Collaboration}, {Brown},
  {Vallenari}, {Prusti}, {de Bruijne}, {Babusiaux}, {Bailer-Jones}, {Biermann},
  {Evans}, {Eyer}, {Jansen}, {Jordi}, {Klioner}, {Lammers}, {Lindegren},
  {Luri}, {Mignard}, {Panem}, {Pourbaix}, {Randich}, {Sartoretti}, {Siddiqui},
  {Soubiran}, {van Leeuwen}, {Walton}, {Arenou}, {Bastian}, {Cropper},
  {Drimmel}, {Katz}, {Lattanzi}, {Bakker}, {Cacciari}, {Casta{\~n}eda},
  {Chaoul}, {Cheek}, {De Angeli}, {Fabricius}, {Guerra}, {Holl}, {Masana},
  {Messineo}, {Mowlavi}, {Nienartowicz}, {Panuzzo}, {Portell}, {Riello},
  {Seabroke}, {Tanga}, {Th{\'e}venin}, {Gracia-Abril}, {Comoretto},
  {Garcia-Reinaldos}, {Teyssier}, {Altmann}, {Andrae}, {Audard},
  {Bellas-Velidis}, {Benson}, {Berthier}, {Blomme}, {Burgess}, {Busso},
  {Carry}, {Cellino}, {Clementini}, {Clotet}, {Creevey}, {Davidson}, {De
  Ridder}, {Delchambre}, {Dell'Oro}, {Ducourant},
  {Fern{\'a}ndez-Hern{\'a}ndez}, {Fouesneau}, {Fr{\'e}mat}, {Galluccio},
  {Garc{\'\i}a-Torres}, {Gonz{\'a}lez-N{\'u}{\~n}ez}, {Gonz{\'a}lez-Vidal},
  {Gosset}, {Guy}, {Halbwachs}, {Hambly}, {Harrison}, {Hern{\'a}ndez},
  {Hestroffer}, {Hodgkin}, {Hutton}, {Jasniewicz}, {Jean-Antoine-Piccolo},
  {Jordan}, {Korn}, {Krone-Martins}, {Lanzafame}, {Lebzelter}, {L{\"o}ffler},
  {Manteiga}, {Marrese}, {Mart{\'\i}n-Fleitas}, {Moitinho}, {Mora}, {Muinonen},
  {Osinde}, {Pancino}, {Pauwels}, {Petit}, {Recio-Blanco}, {Richards},
  {Rimoldini}, {Robin}, {Sarro}, {Siopis}, {Smith}, {Sozzetti}, {S{\"u}veges},
  {Torra}, {van Reeven}, {Abbas}, {Abreu Aramburu}, {Accart}, {Aerts},
  {Altavilla}, {{\'A}lvarez}, {Alvarez}, {Alves}, {Anderson}, {Andrei},
  {Anglada Varela}, {Antiche}, {Antoja}, {Arcay}, {Astraatmadja}, {Bach},
  {Baker}, {Balaguer-N{\'u}{\~n}ez}, {Balm}, {Barache}, {Barata}, {Barbato},
  {Barblan}, {Barklem}, {Barrado}, {Barros}, {Barstow}, {Bartholom{\'e}
  Mu{\~n}oz}, {Bassilana}, {Becciani}, {Bellazzini}, {Berihuete}, {Bertone},
  {Bianchi}, {Bienaym{\'e}}, {Blanco-Cuaresma}, {Boch}, {Boeche}, {Bombrun},
  {Borrachero}, {Bossini}, {Bouquillon}, {Bourda}, {Bragaglia}, {Bramante},
  {Breddels}, {Bressan}, {Brouillet}, {Br{\"u}semeister}, {Brugaletta},
  {Bucciarelli}, {Burlacu}, {Busonero}, {Butkevich}, {Buzzi}, {Caffau},
  {Cancelliere}, {Cannizzaro}, {Cantat-Gaudin}, {Carballo}, {Carlucci},
  {Carrasco}, {Casamiquela}, {Castellani}, {Castro-Ginard}, {Charlot},
  {Chemin}, {Chiavassa}, {Cocozza}, {Costigan}, {Cowell}, {Crifo}, {Crosta},
  {Crowley}, {Cuypers}, {Dafonte}, {Damerdji}, {Dapergolas}, {David}, {David},
  {de Laverny}, {De Luise}, {De March}, {de Martino}, {de Souza}, {de Torres},
  {Debosscher}, {del Pozo}, {Delbo}, {Delgado}, {Delgado}, {Di Matteo},
  {Diakite}, {Diener}, {Distefano}, {Dolding}, {Drazinos}, {Dur{\'a}n},
  {Edvardsson}, {Enke}, {Eriksson}, {Esquej}, {Eynard Bontemps}, {Fabre},
  {Fabrizio}, {Faigler}, {Falc{\~a}o}, {Farr{\`a}s Casas}, {Federici},
  {Fedorets}, {Fernique}, {Figueras}, {Filippi}, {Findeisen}, {Fonti},
  {Fraile}, {Fraser}, {Fr{\'e}zouls}, {Gai}, {Galleti}, {Garabato},
  {Garc{\'\i}a-Sedano}, {Garofalo}, {Garralda}, {Gavel}, {Gavras}, {Gerssen},
  {Geyer}, {Giacobbe}, {Gilmore}, {Girona}, {Giuffrida}, {Glass}, {Gomes},
  {Granvik}, {Gueguen}, {Guerrier}, {Guiraud}, {Guti{\'e}rrez-S{\'a}nchez},
  {Haigron}, {Hatzidimitriou}, {Hauser}, {Haywood}, {Heiter}, {Helmi}, {Heu},
  {Hilger}, {Hobbs}, {Hofmann}, {Holland}, {Huckle}, {Hypki}, {Icardi},
  {Jan{\ss}en}, {Jevardat de Fombelle}, {Jonker}, {Juh{\'a}sz}, {Julbe},
  {Karampelas}, {Kewley}, {Klar}, {Kochoska}, {Kohley}, {Kolenberg},
  {Kontizas}, {Kontizas}, {Koposov}, {Kordopatis}, {Kostrzewa-Rutkowska},
  {Koubsky}, {Lambert}, {Lanza}, {Lasne}, {Lavigne}, {Le Fustec}, {Le
  Poncin-Lafitte}, {Lebreton}, {Leccia}, {Leclerc}, {Lecoeur-Taibi},
  {Lenhardt}, {Leroux}, {Liao}, {Licata}, {Lindstr{\o}m}, {Lister}, {Livanou},
  {Lobel}, {L{\'o}pez}, {Managau}, {Mann}, {Mantelet}, {Marchal}, {Marchant},
  {Marconi}, {Marinoni}, {Marschalk{\'o}}, {Marshall}, {Martino}, {Marton},
  {Mary}, {Massari}, {Matijevi{\v{c}}}, {Mazeh}, {McMillan}, {Messina},
  {Michalik}, {Millar}, {Molina}, {Molinaro}, {Moln{\'a}r}, {Montegriffo},
  {Mor}, {Morbidelli}, {Morel}, {Morris}, {Mulone}, {Muraveva}, {Musella},
  {Nelemans}, {Nicastro}, {Noval}, {O'Mullane}, {Ord{\'e}novic},
  {Ord{\'o}{\~n}ez-Blanco}, {Osborne}, {Pagani}, {Pagano}, {Pailler},
  {Palacin}, {Palaversa}, {Panahi}, {Pawlak}, {Piersimoni}, {Pineau}, {Plachy},
  {Plum}, {Poggio}, {Poujoulet}, {Pr{\v{s}}a}, {Pulone}, {Racero}, {Ragaini},
  {Rambaux}, {Ramos-Lerate}, {Regibo}, {Reyl{\'e}}, {Riclet}, {Ripepi}, {Riva},
  {Rivard}, {Rixon}, {Roegiers}, {Roelens}, {Romero-G{\'o}mez}, {Rowell},
  {Royer}, {Ruiz-Dern}, {Sadowski}, {Sagrist{\`a} Sell{\'e}s}, {Sahlmann},
  {Salgado}, {Salguero}, {Sanna}, {Santana-Ros}, {Sarasso}, {Savietto},
  {Schultheis}, {Sciacca}, {Segol}, {Segovia}, {S{\'e}gransan}, {Shih},
  {Siltala}, {Silva}, {Smart}, {Smith}, {Solano}, {Solitro}, {Sordo}, {Soria
  Nieto}, {Souchay}, {Spagna}, {Spoto}, {Stampa}, {Steele},
  {Steidelm{\"u}ller}, {Stephenson}, {Stoev}, {Suess}, {Surdej}, {Szabados},
  {Szegedi-Elek}, {Tapiador}, {Taris}, {Tauran}, {Taylor}, {Teixeira},
  {Terrett}, {Teyssand ier}, {Thuillot}, {Titarenko}, {Torra Clotet}, {Turon},
  {Ulla}, {Utrilla}, {Uzzi}, {Vaillant}, {Valentini}, {Valette}, {van Elteren},
  {Van Hemelryck}, {van Leeuwen}, {Vaschetto}, {Vecchiato}, {Veljanoski},
  {Viala}, {Vicente}, {Vogt}, {von Essen}, {Voss}, {Votruba}, {Voutsinas},
  {Walmsley}, {Weiler}, {Wertz}, {Wevers}, {Wyrzykowski}, {Yoldas},
  {{\v{Z}}erjal}, {Ziaeepour}, {Zorec}, {Zschocke}, {Zucker}, {Zurbach}, \&
  {Zwitter}}]{Gaia2018}
{Gaia Collaboration}, {Brown}, A.~G.~A., {Vallenari}, A., {et~al.} 2018, \aap,
  616, A1, \dodoi{10.1051/0004-6361/201833051}

\bibitem[{{GRAVITY Collaboration} {et~al.}(2020){GRAVITY Collaboration},
  {Nowak, M.}, {Lacour, S.}, {Molli\`ere, P.}, {Wang, J.}, {Charnay, B.}, {van
  Dishoeck, E. F.}, {Abuter, R.}, {Amorim, A.}, {Berger, J. P.}, {Beust, H.},
  {Bonnefoy, M.}, {Bonnet, H.}, {Brandner, W.}, {Buron, A.}, {Cantalloube, F.},
  {Collin, C.}, {Chapron, F.}, {Cl\'enet, Y.}, {Coud\'e du Foresto, V.}, {de
  Zeeuw, P. T.}, {Dembet, R.}, {Dexter, J.}, {Duvert, G.}, {Eckart, A.},
  {Eisenhauer, F.}, {F\"orster Schreiber, N. M.}, {F\'edou, P.}, {Garcia Lopez,
  R.}, {Gao, F.}, {Gendron, E.}, {Genzel, R.}, {Gillessen, S.}, {Hau\ss{}mann,
  F.}, {Henning, T.}, {Hippler, S.}, {Hubert, Z.}, {Jocou, L.}, {Kervella, P.},
  {Lagrange, A.-M.}, {Lapeyr\`ere, V.}, {Le Bouquin, J.-B.}, {L\'ena, P.},
  {Maire, A.-L.}, {Ott, T.}, {Paumard, T.}, {Paladini, C.}, {Perraut, K.},
  {Perrin, G.}, {Pueyo, L.}, {Pfuhl, O.}, {Rabien, S.}, {Rau, C.},
  {Rodr\'{\i}guez-Coira, G.}, {Rousset, G.}, {Scheithauer, S.}, {Shangguan,
  J.}, {Straub, O.}, {Straubmeier, C.}, {Sturm, E.}, {Tacconi, L. J.},
  {Vincent, F.}, {Widmann, F.}, {Wieprecht, E.}, {Wiezorrek, E.}, {Woillez,
  J.}, {Yazici, S.}, \& {Ziegler, D.}}]{Nowak_etal2020}
{GRAVITY Collaboration}, {Nowak, M.}, {Lacour, S.}, {et~al.} 2020, A\&A, 633,
  A110, \dodoi{10.1051/0004-6361/201936898}

\bibitem[{{Hatzes} {et~al.}(2000){Hatzes}, {Cochran}, {McArthur}, {Baliunas},
  {Walker}, {Campbell}, {Irwin}, {Yang}, {K{\"u}rster}, {Endl}, {Els},
  {Butler}, \& {Marcy}}]{Hatzes2000}
{Hatzes}, A.~P., {Cochran}, W.~D., {McArthur}, B., {et~al.} 2000, \apjl, 544,
  L145, \dodoi{10.1086/317319}

\bibitem[{{Hinkley} {et~al.}(2021){Hinkley}, {Matthews}, {Lefevre}, {Lestrade},
  {Kennedy}, {Mawet}, {Stapelfeldt}, {Ray}, {Mamajek}, {Bowler}, {Wilner},
  {Williams}, {Ansdell}, {Wyatt}, {Lau}, {Phillips}, {Fernandez}, {Gagn{\'e}},
  {Bubb}, {Sutlieff}, {Wilson}, {Matthews}, {Ngo}, {Piskorz}, {Crepp},
  {Gonzalez}, {Mann}, \& {Mace}}]{Hinkley2021}
{Hinkley}, S., {Matthews}, E.~C., {Lefevre}, C., {et~al.} 2021, \apj, 912, 115,
  \dodoi{10.3847/1538-4357/abec6e}

\bibitem[{{Hoeijmakers} {et~al.}(2018){Hoeijmakers}, {Schwarz}, {Snellen}, {de
  Kok}, {Bonnefoy}, {Chauvin}, {Lagrange}, \& {Girard}}]{Hoeijmakers2018}
{Hoeijmakers}, H.~J., {Schwarz}, H., {Snellen}, I.~A.~G., {et~al.} 2018, \aap,
  617, A144, \dodoi{10.1051/0004-6361/201832902}

\bibitem[{Hou {et~al.}(2012)Hou, Goodman, Hogg, Weare, \& Schwab}]{Hou2012}
Hou, F., Goodman, J., Hogg, D.~W., Weare, J., \& Schwab, C. 2012, The
  Astrophysical Journal, 745, 198, \dodoi{10.1088/0004-637x/745/2/198}

\bibitem[{{Houll{\'e}} {et~al.}(2021){Houll{\'e}}, {Vigan}, {Carlotti},
  {Choquet}, {Cantalloube}, {Phillips}, {Sauvage}, {Schwartz}, {Otten},
  {Baraffe}, {Emsenhuber}, \& {Mordasini}}]{Houlle2021}
{Houll{\'e}}, M., {Vigan}, A., {Carlotti}, A., {et~al.} 2021, arXiv e-prints,
  arXiv:2104.11251.
\newblock \doarXiv{2104.11251}

\bibitem[{{Howard} {et~al.}(2010){Howard}, {Johnson}, {Marcy}, {Fischer},
  {Wright}, {Bernat}, {Henry}, {Peek}, {Isaacson}, {Apps}, {Endl}, {Cochran},
  {Valenti}, {Anderson}, \& {Piskunov}}]{Howard2010}
{Howard}, A.~W., {Johnson}, J.~A., {Marcy}, G.~W., {et~al.} 2010, \apj, 721,
  1467, \dodoi{10.1088/0004-637X/721/2/1467}

\bibitem[{{Huby} {et~al.}(2015){Huby}, {Baudoz}, {Mawet}, \&
  {Absil}}]{Huby2015}
{Huby}, E., {Baudoz}, P., {Mawet}, D., \& {Absil}, O. 2015, \aap, 584, A74,
  \dodoi{10.1051/0004-6361/201527102}

\bibitem[{{Huby} {et~al.}(2017){Huby}, {Bottom}, {Femenia}, {Ngo}, {Mawet},
  {Serabyn}, \& {Absil}}]{Huby2017}
{Huby}, E., {Bottom}, M., {Femenia}, B., {et~al.} 2017, \aap, 600, A46,
  \dodoi{10.1051/0004-6361/201630232}

\bibitem[{Hunter(2007)}]{Matplotlib}
Hunter, J.~D. 2007, Computing in Science Engineering, 9, 90,
  \dodoi{10.1109/MCSE.2007.55}

\bibitem[{{Janson} {et~al.}(2015){Janson}, {Quanz}, {Carson}, {Thalmann},
  {Lafreni{\`e}re}, \& {Amara}}]{Janson2015}
{Janson}, M., {Quanz}, S.~P., {Carson}, J.~C., {et~al.} 2015, \aap, 574, A120,
  \dodoi{10.1051/0004-6361/201424944}

\bibitem[{Jones {et~al.}(2001)Jones, Oliphant, \& Peterson}]{Jones2001}
Jones, E., Oliphant, T., \& Peterson, P. 2001

\bibitem[{{Kasper} {et~al.}(2019){Kasper}, {Arsenault}, {K{\"a}ufl}, {Jakob},
  {Leveratto}, {Zins}, {Pantin}, {Duhoux}, {Riquelme}, {Kirchbauer}, {Kolb},
  {Pathak}, {Siebenmorgen}, {Soenke}, {Fuenteseca}, {Sterzik}, {Ageorges},
  {Gutruf}, {Kampf}, {Reutlinger}, {Absil}, {Delacroix}, {Maire}, {Huby},
  {Guyon}, {Klupar}, {Mawet}, {Ruane}, {Karlsson}, {Dohlen}, {Vigan},
  {N'Diaye}, {Quanz}, \& {Carlotti}}]{Kasper2019}
{Kasper}, M., {Arsenault}, R., {K{\"a}ufl}, U., {et~al.} 2019, The Messenger,
  178, 5, \dodoi{10.18727/0722-6691/5163}

\bibitem[{{Keenan} \& {McNeil}(1989)}]{Keenan1989}
{Keenan}, P.~C., \& {McNeil}, R.~C. 1989, \apjs, 71, 245,
  \dodoi{10.1086/191373}

\bibitem[{{Kervella} {et~al.}(2019){Kervella}, {Arenou, Fr\'ed\'eric},
  {Mignard, Fran\c{c}ois}, \& {Th\'evenin, Fr\'ed\'eric}}]{Kervella2019}
{Kervella}, P., {Arenou, Fr\'ed\'eric}, {Mignard, Fran\c{c}ois}, \&
  {Th\'evenin, Fr\'ed\'eric}. 2019, A\&A, 623, A72,
  \dodoi{10.1051/0004-6361/201834371}

\bibitem[{{Konopacky} {et~al.}(2013){Konopacky}, {Barman}, {Macintosh}, \&
  {Marois}}]{Konopacky2013}
{Konopacky}, Q.~M., {Barman}, T.~S., {Macintosh}, B.~A., \& {Marois}, C. 2013,
  Science, 339, 1398, \dodoi{10.1126/science.1232003}

\bibitem[{{Lindegren} {et~al.}(2018){Lindegren}, {Hern{\'a}ndez}, {Bombrun},
  {Klioner}, {Bastian}, {Ramos-Lerate}, {de Torres}, {Steidelm{\"u}ller},
  {Stephenson}, {Hobbs}, {Lammers}, {Biermann}, {Geyer}, {Hilger}, {Michalik},
  {Stampa}, {McMillan}, {Casta{\~n}eda}, {Clotet}, {Comoretto}, {Davidson},
  {Fabricius}, {Gracia}, {Hambly}, {Hutton}, {Mora}, {Portell}, {van Leeuwen},
  {Abbas}, {Abreu}, {Altmann}, {Andrei}, {Anglada}, {Balaguer-N{\'u}{\~n}ez},
  {Barache}, {Becciani}, {Bertone}, {Bianchi}, {Bouquillon}, {Bourda},
  {Br{\"u}semeister}, {Bucciarelli}, {Busonero}, {Buzzi}, {Cancelliere},
  {Carlucci}, {Charlot}, {Cheek}, {Crosta}, {Crowley}, {de Bruijne}, {de
  Felice}, {Drimmel}, {Esquej}, {Fienga}, {Fraile}, {Gai}, {Garralda},
  {Gonz{\'a}lez-Vidal}, {Guerra}, {Hauser}, {Hofmann}, {Holl}, {Jordan},
  {Lattanzi}, {Lenhardt}, {Liao}, {Licata}, {Lister}, {L{\"o}ffler},
  {Marchant}, {Martin-Fleitas}, {Messineo}, {Mignard}, {Morbidelli}, {Poggio},
  {Riva}, {Rowell}, {Salguero}, {Sarasso}, {Sciacca}, {Siddiqui}, {Smart},
  {Spagna}, {Steele}, {Taris}, {Torra}, {van Elteren}, {van Reeven}, \&
  {Vecchiato}}]{Lindegren2018}
{Lindegren}, L., {Hern{\'a}ndez}, J., {Bombrun}, A., {et~al.} 2018, \aap, 616,
  A2, \dodoi{10.1051/0004-6361/201832727}

\bibitem[{{L{\'o}pez-Morales} {et~al.}(2016){L{\'o}pez-Morales}, {Haywood},
  {Coughlin}, {Zeng}, {Buchhave}, {Giles}, {Affer}, {Bonomo}, {Charbonneau},
  {Collier Cameron}, {Consentino}, {Dressing}, {Dumusque}, {Figueira},
  {Fiorenzano}, {Harutyunyan}, {Johnson}, {Latham}, {Lopez}, {Lovis},
  {Malavolta}, {Mayor}, {Micela}, {Molinari}, {Mortier}, {Motalebi},
  {Nascimbeni}, {Pepe}, {Phillips}, {Piotto}, {Pollacco}, {Queloz}, {Rice},
  {Sasselov}, {Segransan}, {Sozzetti}, {Udry}, {Vanderburg}, \&
  {Watson}}]{Lopez-Morales2016}
{L{\'o}pez-Morales}, M., {Haywood}, R.~D., {Coughlin}, J.~L., {et~al.} 2016,
  \aj, 152, 204, \dodoi{10.3847/0004-6256/152/6/204}

\bibitem[{{MacGregor} {et~al.}(2015){MacGregor}, {Wilner}, {Andrews},
  {Lestrade}, \& {Maddison}}]{MacGregor2015}
{MacGregor}, M.~A., {Wilner}, D.~J., {Andrews}, S.~M., {Lestrade}, J.-F., \&
  {Maddison}, S. 2015, \apj, 809, 47, \dodoi{10.1088/0004-637X/809/1/47}

\bibitem[{{Mamajek} \& {Hillenbrand}(2008)}]{Mamajek2008}
{Mamajek}, E.~E., \& {Hillenbrand}, L.~A. 2008, \apj, 687, 1264,
  \dodoi{10.1086/591785}

\bibitem[{Marley {et~al.}(2018)Marley, Saumon, Morley, \&
  Fortney}]{Marley2018sonora}
Marley, M., Saumon, D., Morley, C., \& Fortney, J. 2018, {Sonora 2018:
  Cloud-free, solar composition, solar C/O substellar atmosphere models and
  spectra}, nc\_m+0.0\_co1.0\_v1.0,  Zenodo, \dodoi{10.5281/zenodo.1309035}

\bibitem[{{Marois} {et~al.}(2008){Marois}, {Lafreni{\`e}re}, {Macintosh}, \&
  {Doyon}}]{Marois2008}
{Marois}, C., {Lafreni{\`e}re}, D., {Macintosh}, B., \& {Doyon}, R. 2008, \apj,
  673, 647, \dodoi{10.1086/523839}

\bibitem[{{Mawet} {et~al.}(2019){Mawet}, {Hirsch}, {Lee}, {Ruffio}, {Bottom},
  {Fulton}, {Absil}, {Beichman}, {Bowler}, {Bryan}, {Choquet}, {Ciardi},
  {Christiaens}, {Defr{\`e}re}, {Gomez Gonzalez}, {Howard}, {Huby}, {Isaacson},
  {Jensen-Clem}, {Kosiarek}, {Marcy}, {Meshkat}, {Petigura}, {Reggiani},
  {Ruane}, {Serabyn}, {Sinukoff}, {Wang}, {Weiss}, \& {Ygouf}}]{Mawet2019}
{Mawet}, D., {Hirsch}, L., {Lee}, E.~J., {et~al.} 2019, \aj, 157, 33,
  \dodoi{10.3847/1538-3881/aaef8a}

\bibitem[{{Nielsen} {et~al.}(2020){Nielsen}, {De Rosa}, {Wang}, {Sahlmann},
  {Kalas}, {Duch{\^e}ne}, {Rameau}, {Marley}, {Saumon}, {Macintosh},
  {Millar-Blanchaer}, {Nguyen}, {Ammons}, {Bailey}, {Barman}, {Bulger},
  {Chilcote}, {Cotten}, {Doyon}, {Esposito}, {Fitzgerald}, {Follette},
  {Gerard}, {Goodsell}, {Graham}, {Greenbaum}, {Hibon}, {Hung}, {Ingraham},
  {Konopacky}, {Larkin}, {Maire}, {Marchis}, {Marois}, {Metchev},
  {Oppenheimer}, {Palmer}, {Patience}, {Perrin}, {Poyneer}, {Pueyo}, {Rajan},
  {Rantakyr{\"o}}, {Ruffio}, {Savransky}, {Schneider}, {Sivaramakrishnan},
  {Song}, {Soummer}, {Thomas}, {Wallace}, {Ward-Duong}, {Wiktorowicz}, \&
  {Wolff}}]{Nielsen2020}
{Nielsen}, E.~L., {De Rosa}, R.~J., {Wang}, J.~J., {et~al.} 2020, \aj, 159, 71,
  \dodoi{10.3847/1538-3881/ab5b92}

\bibitem[{{Nowak} {et~al.}(2020){Nowak}, {Lacour}, {Lagrange}, {Rubini},
  {Wang}, {Stolker}, {Abuter}, {Amorim}, {Asensio-Torres}, {Baub{\"o}ck},
  {Benisty}, {Berger}, {Beust}, {Blunt}, {Boccaletti}, {Bonnefoy}, {Bonnet},
  {Brandner}, {Cantalloube}, {Charnay}, {Choquet}, {Christiaens}, {Cl{\'e}net},
  {Coud{\'e} Du Foresto}, {Cridland}, {de Zeeuw}, {Dembet}, {Dexter},
  {Drescher}, {Duvert}, {Eckart}, {Eisenhauer}, {Gao}, {Garcia}, {Garcia
  Lopez}, {Gardner}, {Gendron}, {Genzel}, {Gillessen}, {Girard}, {Grandjean},
  {Haubois}, {Hei{\ss}el}, {Henning}, {Hinkley}, {Hippler}, {Horrobin},
  {Houll{\'e}}, {Hubert}, {Jim{\'e}nez-Rosales}, {Jocou}, {Kammerer},
  {Kervella}, {Keppler}, {Kreidberg}, {Kulikauskas}, {Lapeyr{\`e}re}, {Le
  Bouquin}, {L{\'e}na}, {M{\'e}rand}, {Maire}, {Molli{\`e}re}, {Monnier},
  {Mouillet}, {M{\"u}ller}, {Nasedkin}, {Ott}, {Otten}, {Paumard}, {Paladini},
  {Perraut}, {Perrin}, {Pueyo}, {Pfuhl}, {Rameau}, {Rodet},
  {Rodr{\'\i}guez-Coira}, {Rousset}, {Scheithauer}, {Shangguan}, {Stadler},
  {Straub}, {Straubmeier}, {Sturm}, {Tacconi}, {van Dishoeck}, {Vigan},
  {Vincent}, {von Fellenberg}, {Ward-Duong}, {Widmann}, {Wieprecht},
  {Wiezorrek}, {Woillez}, \& {Gravity Collaboration}}]{Nowak2020}
{Nowak}, M., {Lacour}, S., {Lagrange}, A.~M., {et~al.} 2020, \aap, 642, L2,
  \dodoi{10.1051/0004-6361/202039039}

\bibitem[{{Pathak} {et~al.}(2021){Pathak}, {Petit dit de la Roche}, {Kasper},
  {Sterzik}, {Absil}, {Boehle}, {Feng}, {Ivanov}, {Janson}, {Jones}, {Kaufer},
  {K{\"a}ufl}, {Maire}, {Meyer}, {Pantin}, {Siebenmorgen}, {van den Ancker}, \&
  {Viswanath}}]{Pathak2021}
{Pathak}, P., {Petit dit de la Roche}, D.~J.~M., {Kasper}, M., {et~al.} 2021,
  arXiv e-prints, arXiv:2104.13032.
\newblock \doarXiv{2104.13032}

\bibitem[{Perrin {et~al.}(2018)Perrin, Pueyo, Gorkom, Brooks, Rajan, Girard, \&
  Lajoie}]{Perrin2018}
Perrin, M.~D., Pueyo, L., Gorkom, K.~V., {et~al.} 2018, in Space Telescopes and
  Instrumentation 2018: Optical, Infrared, and Millimeter Wave, ed. M.~Lystrup,
  H.~A. MacEwen, G.~G. Fazio, N.~Batalha, N.~Siegler, \& E.~C. Tong, Vol.
  10698, International Society for Optics and Photonics (SPIE), 98 -- 113,
  \dodoi{10.1117/12.2313552}

\bibitem[{{Petit dit de la Roche} {et~al.}(2018){Petit dit de la Roche},
  {Hoeijmakers}, \& {Snellen}}]{PetitditdelaRoche2018}
{Petit dit de la Roche}, D.~J.~M., {Hoeijmakers}, H.~J., \& {Snellen}, I.~A.~G.
  2018, \aap, 616, A146, \dodoi{10.1051/0004-6361/201833384}

\bibitem[{{Quillen} \& {Faber}(2006)}]{Quillen2006}
{Quillen}, A.~C., \& {Faber}, P. 2006, \mnras, 373, 1245,
  \dodoi{10.1111/j.1365-2966.2006.11122.x}

\bibitem[{{Radovan} {et~al.}(2014){Radovan}, {Lanclos}, {Holden}, {Kibrick},
  {Allen}, {Deich}, {Rivera}, {Burt}, {Fulton}, {Butler}, \&
  {Vogt}}]{Radovan2014}
{Radovan}, M.~V., {Lanclos}, K., {Holden}, B.~P., {et~al.} 2014, in Society of
  Photo-Optical Instrumentation Engineers (SPIE) Conference Series, Vol. 9145,
  Ground-based and Airborne Telescopes V, ed. L.~M. {Stepp}, R.~{Gilmozzi}, \&
  H.~J. {Hall}, 91452B, \dodoi{10.1117/12.2057310}

\bibitem[{Rajpaul {et~al.}(2015)Rajpaul, Aigrain, Osborne, Reece, \&
  Roberts}]{Rajpaul2015}
Rajpaul, V., Aigrain, S., Osborne, M.~A., Reece, S., \& Roberts, S. 2015,
  Monthly Notices of the Royal Astronomical Society, 452, 2269,
  \dodoi{10.1093/mnras/stv1428}

\bibitem[{{Ruffio} {et~al.}(2017){Ruffio}, {Macintosh}, {Wang}, {Pueyo},
  {Nielsen}, {De Rosa}, {Czekala}, {Marley}, {Arriaga}, {Bailey}, {Barman},
  {Bulger}, {Chilcote}, {Cotten}, {Doyon}, {Duch{\^e}ne}, {Fitzgerald},
  {Follette}, {Gerard}, {Goodsell}, {Graham}, {Greenbaum}, {Hibon}, {Hung},
  {Ingraham}, {Kalas}, {Konopacky}, {Larkin}, {Maire}, {Marchis}, {Marois},
  {Metchev}, {Millar-Blanchaer}, {Morzinski}, {Oppenheimer}, {Palmer},
  {Patience}, {Perrin}, {Poyneer}, {Rajan}, {Rameau}, {Rantakyr{\"o}},
  {Savransky}, {Schneider}, {Sivaramakrishnan}, {Song}, {Soummer}, {Thomas},
  {Wallace}, {Ward-Duong}, {Wiktorowicz}, \& {Wolff}}]{Ruffio2017}
{Ruffio}, J.-B., {Macintosh}, B., {Wang}, J.~J., {et~al.} 2017, ApJ, 842, 14,
  \dodoi{10.3847/1538-4357/aa72dd}

\bibitem[{{Ruffio} {et~al.}(2018){Ruffio}, {Mawet}, {Czekala}, {Macintosh}, {De
  Rosa}, {Ruane}, {Bottom}, {Pueyo}, {Wang}, {Hirsch}, {Zhu}, \&
  {Nielsen}}]{Ruffio2018}
{Ruffio}, J.-B., {Mawet}, D., {Czekala}, I., {et~al.} 2018, \aj, 156, 196,
  \dodoi{10.3847/1538-3881/aade95}

\bibitem[{{Ruffio} {et~al.}(2019){Ruffio}, {Macintosh}, {Konopacky}, {Barman},
  {De Rosa}, {Wang}, {Wilcomb}, {Czekala}, \& {Marois}}]{Ruffio2019}
{Ruffio}, J.-B., {Macintosh}, B., {Konopacky}, Q.~M., {et~al.} 2019, \aj, 158,
  200, \dodoi{10.3847/1538-3881/ab4594}

\bibitem[{{Serabyn} {et~al.}(2017){Serabyn}, {Huby}, {Matthews}, {Mawet},
  {Absil}, {Femenia}, {Wizinowich}, {Karlsson}, {Bottom}, {Campbell},
  {Carlomagno}, {Defr{\`e}re}, {Delacroix}, {Forsberg}, {Gomez Gonzalez},
  {Habraken}, {Jolivet}, {Liewer}, {Lilley}, {Piron}, {Reggiani}, {Surdej},
  {Tran}, {Vargas Catal{\'a}n}, \& {Wertz}}]{Serabyn2017}
{Serabyn}, E., {Huby}, E., {Matthews}, K., {et~al.} 2017, \aj, 153, 43,
  \dodoi{10.3847/1538-3881/153/1/43}

\bibitem[{Soummer {et~al.}(2012)Soummer, Pueyo, \& Larkin}]{Soummer2012}
Soummer, R., Pueyo, L., \& Larkin, J. 2012, The Astrophysical Journal, 755,
  L28, \dodoi{10.1088/2041-8205/755/2/l28}

\bibitem[{{Su} {et~al.}(2017){Su}, {De Buizer}, {Rieke}, {Krivov}, {L{\"o}hne},
  {Marengo}, {Stapelfeldt}, {Ballering}, \& {Vacca}}]{Su2017}
{Su}, K.~Y.~L., {De Buizer}, J.~M., {Rieke}, G.~H., {et~al.} 2017, \aj, 153,
  226, \dodoi{10.3847/1538-3881/aa696b}

\bibitem[{{van Leeuwen}(2007)}]{vanLeeuwen2007}
{van Leeuwen}, F. 2007, {Hipparcos, the New Reduction of the Raw Data}, Vol.
  350, \dodoi{10.1007/978-1-4020-6342-8}

\bibitem[{{Vogt} {et~al.}(1994){Vogt}, {Allen}, {Bigelow}, {Bresee}, {Brown},
  {Cantrall}, {Conrad}, {Couture}, {Delaney}, {Epps}, {Hilyard}, {Hilyard},
  {Horn}, {Jern}, {Kanto}, {Keane}, {Kibrick}, {Lewis}, {Osborne},
  {Pardeilhan}, {Pfister}, {Ricketts}, {Robinson}, {Stover}, {Tucker}, {Ward},
  \& {Wei}}]{Vogt1994}
{Vogt}, S.~S., {Allen}, S.~L., {Bigelow}, B.~C., {et~al.} 1994, in Society of
  Photo-Optical Instrumentation Engineers (SPIE) Conference Series, Vol. 2198,
  Instrumentation in Astronomy VIII, ed. D.~L. {Crawford} \& E.~R. {Craine},
  362, \dodoi{10.1117/12.176725}

\bibitem[{{Vogt} {et~al.}(2014){Vogt}, {Radovan}, {Kibrick}, {Butler},
  {Alcott}, {Allen}, {Arriagada}, {Bolte}, {Burt}, {Cabak}, {Chloros},
  {Cowley}, {Deich}, {Dupraw}, {Earthman}, {Epps}, {Faber}, {Fischer}, {Gates},
  {Hilyard}, {Holden}, {Johnston}, {Keiser}, {Kanto}, {Katsuki}, {Laiterman},
  {Lanclos}, {Laughlin}, {Lewis}, {Lockwood}, {Lynam}, {Marcy}, {McLean},
  {Miller}, {Misch}, {Peck}, {Pfister}, {Phillips}, {Rivera}, {Sand ford},
  {Saylor}, {Stover}, {Thompson}, {Walp}, {Ward}, {Wareham}, {Wei}, \&
  {Wright}}]{Vogt2014}
{Vogt}, S.~S., {Radovan}, M., {Kibrick}, R., {et~al.} 2014, \pasp, 126, 359,
  \dodoi{10.1086/676120}

\bibitem[{{Wang} {et~al.}(2015){Wang}, {Ruffio}, {De Rosa}, {Aguilar}, {Wolff},
  \& {Pueyo}}]{Wang2015}
{Wang}, J.~J., {Ruffio}, J.-B., {De Rosa}, R.~J., {et~al.} 2015, {pyKLIP: PSF
  Subtraction for Exoplanets and Disks}, Astrophysics Source Code Library.
\newblock \doeprint{1506.001}

\bibitem[{Wang {et~al.}(2021)Wang, Vigan, Lacour, Nowak, Stolker, Rosa,
  Ginzburg, Gao, Abuter, Amorim, Asensio-Torres, Baub{\"o}ck, Benisty, Berger,
  Beust, Beuzit, Blunt, Boccaletti, Bohn, Bonnefoy, Bonnet, Brandner,
  Cantalloube, Caselli, Charnay, Chauvin, Choquet, Christiaens, Cl{\'{e}}net,
  du~Foresto, Cridland, de~Zeeuw, Dembet, Dexter, Drescher, Duvert, Eckart,
  Eisenhauer, Facchini, Gao, Garcia, Lopez, Gardner, Gendron, Genzel,
  Gillessen, Girard, Haubois, Hei{\ss}el, Henning, Hinkley, Hippler, Horrobin,
  Houll{\'{e}}, Hubert, Jim{\'{e}}nez-Rosales, Jocou, Kammerer, Keppler,
  Kervella, Meyer, Kreidberg, Lagrange, Lapeyr{\`{e}}re, Bouquin, L{\'{e}}na,
  Lutz, Maire, M{\'{e}}nard, M{\'{e}}rand, Molli{\`{e}}re, Monnier, Mouillet,
  M{\"u}ller, Nasedkin, Ott, Otten, Paladini, Paumard, Perraut, Perrin, Pfuhl,
  Pueyo, Rameau, Rodet, Rodr{\'{\i}}guez-Coira, Rousset, Scheithauer,
  Shangguan, Shimizu, Stadler, Straub, Straubmeier, Sturm, Tacconi, van
  Dishoeck, Vincent, von Fellenberg, Ward-Duong, Widmann, Wieprecht, Wiezorrek,
  \& Woillez}]{Wang2021}
Wang, J.~J., Vigan, A., Lacour, S., {et~al.} 2021, The Astronomical Journal,
  161, 148, \dodoi{10.3847/1538-3881/abdb2d}

\bibitem[{{Wilcomb} {et~al.}(2020){Wilcomb}, {Konopacky}, {Barman}, {Theissen},
  {Ruffio}, {Brock}, {Macintosh}, \& {Marois}}]{Wilcomb2020}
{Wilcomb}, K.~K., {Konopacky}, Q.~M., {Barman}, T.~S., {et~al.} 2020, \aj, 160,
  207, \dodoi{10.3847/1538-3881/abb9b1}

\bibitem[{{Wisdom}(1980)}]{Wisdom1980}
{Wisdom}, J. 1980, \aj, 85, 1122, \dodoi{10.1086/112778}

\bibitem[{{Xuan} {et~al.}(2018){Xuan}, {Mawet}, {Ngo}, {Ruane}, {Bailey},
  {Choquet}, {Absil}, {Alvarez}, {Bryan}, {Cook}, {Femen{\'\i}a Castell{\'a}},
  {Gomez Gonzalez}, {Huby}, {Knutson}, {Matthews}, {Ragland}, {Serabyn}, \&
  {Zawol}}]{Xuan18}
{Xuan}, W.~J., {Mawet}, D., {Ngo}, H., {et~al.} 2018, \aj, 156, 156,
  \dodoi{10.3847/1538-3881/aadae6}

\end{thebibliography}
\bibliographystyle{aasjournal}

\end{document}